\documentclass[12pt, a4paper, twoside, reqno]{amsart}

\usepackage[english]{babel}

\usepackage{geometry}

\usepackage[noadjust]{cite}

\usepackage{amssymb}
\usepackage{amsfonts}
\usepackage{amsmath}
\usepackage{latexsym}

\usepackage{scalerel}

\usepackage{caption}

\usepackage{url}
\usepackage{hyperref}
\usepackage{amsthm}

\usepackage{tabularx}

\usepackage{enumitem}
\makeatletter
\newcommand*\bigcdot{\mathpalette\bigcdot@{.75}}
\newcommand*\bigcdot@[2]{\mathbin{\vcenter{\hbox{\scalebox{#2}{$\m@th#1\bullet$}}}}}
\makeatother

\hypersetup{
    colorlinks = true,
    citecolor = black,
    filecolor = black,
    linkcolor = black,
    urlcolor = black
}
\newcommand{\removeEdgesSet}[2]{{#1}^{\lnot{#2}}}

\newcommand{\intInterval}[2]{[\![#1,#2]\!]}

\def\BoolSet{\textsc{Bool}}
\def\True{\textsc{True}}
\def\False{\textsc{False}}

\def\LCVP{LCVP}
\def\LCVPfull{locally checkable vertex partitioning}

\def\LCC{LCC}
\def\LCCfull{local condition composition}

\def\ECC{ECC}
\def\ECCfull{edge condition composition}

\newcommand{\rLC}[1]{$#1$-locally checkable}
\newcommand\rLCproblem[1]{{\rLC{#1}} problem}
\newcommand\rLCproblems[1]{{\rLC{#1}} problems}

\def\genDomSym{\lambda}

\newcommand{\genDomDomain}[1]{\mathcal{D}_{#1}}
\def\LabelSet{\textsc{Labels}}
\def\ColorSet{\textsc{Colors}}
\def\WeightSet{\textsc{Weights}}

\def\wlesseq{\preceq}

\def\minWeights{\min}

\def\Error{\textsc{Error}}

\def\weightsSum{\oplus}
\def\bigweightsSum{\bigoplus}

\def\weightsNeutral{e_{\weightsSum}}

\def\weightsFun{\omega}
\def\globalPropSet{\Pi}
\newcommand{\NeighborhoodSet}[2]{\mathcal{N}_{{#1}, {#2}}}

\def\neighborhoodSumSym{\boxplus}
\def\bigneighborhoodSumSym{\bigboxplus}
\newcommand{\neighborhoodSum}[2]{\neighborhoodSumSym^{{#1}, {#2}}}

\def\toNeighborhoodSym{newN}
\newcommand{\toNeighborhood}[2]{{\toNeighborhoodSym}_{{#1}, {#2}}}

\newcommand{\emptyNeighborhood}[2]{e_{{#1}, {#2}}}

\def\pnsubg{\textsc{ns}}

\def\pnPolynomial{polynomial}

\def\pnConstant{constant}
\def\checkFuns{\check{c}}
\def\notInSeq{\perp}

\newcommand{\defProb}[5]{
	\noindent{\renewcommand{\arraystretch}{1}\setlength{\tabcolsep}{2pt}
		\begin{tabularx}{\textwidth}{lX}
			\multicolumn{2}{X}{\sc #1}			\\ % Title
			\emph{{#2}:}	& \multicolumn{1}{X}{#3}	\\ % Input
			\emph{{#4}:}	& \multicolumn{1}{X}{#5}	\\ % Question
		\end{tabularx}
	}
	\smallskip
}

\newcommand{\defProbEnglish}[3]{
	\defProb{#1}{Instance}{#2}{Question}{#3}
}
\newcommand{\setst}[2]{\{ #1 : #2 \}}

\newcommand{\seq}[1]{\left( #1 \right)}

\DeclareMathOperator*{\bigboxplus}{\scalerel*{\boxplus}{\sum}}

\newcommand{\quantifierFormula}[3]{#1 #2 .\, #3}

\newcommand{\forallFormula}[2]{\quantifierFormula{\forall}{#1}{#2}}

\newcommand{\existsFormula}[2]{\quantifierFormula{\exists}{#1}{#2}}

\newcommand{\existsUniqueFormula}[2]{\quantifierFormula{\exists !}{#1}{#2}}

\newtheoremstyle{TheoremNum}
		{8pt}{8pt}								% space between body and thm
		{\itshape}								% thm body font
		{}										% indent amount (empty = no indent)
		{\bfseries}								% thm head font
		{.}										% punctuation after thm head
		{5pt plus 1pt minus 1pt}					% space after thm head
		{\thmname{#1} \thmnote{\bfseries #3}}	% thm head spec

\theoremstyle{TheoremNum}
\newtheorem{theoremRep}{Theorem}
\newtheorem{corollaryRep}{Corollary}

\theoremstyle{plain}
\newtheorem{lemma}{Lemma}[subsection]
\newtheorem{proposition}[lemma]{Proposition}
\newtheorem{theorem}[lemma]{Theorem}

\newtheorem{corollary}[lemma]{Corollary}

\theoremstyle{remark}

\newtheorem{remark}[lemma]{Remark}

\theoremstyle{definition}
\newtheorem{definition}[lemma]{Definition}

\def\THMmainTheorem{
Let $\mathcal{F}$ be a family of graphs of bounded treewidth. Consider a family of instances of a {\rLCproblem{1}} with a {\pnPolynomial} partial neighborhood system, where
\begin{itemize}
\item $G \in \mathcal{F}$,
\item $\mathcal{C} = \max\setst{|L_v|}{v \in V(G)}$ is polynomial in the input size, and
\item the functions $\weightsSum$ and $\minWeights$ can be computed in polynomial time.
\end{itemize}

Then there exists an algorithm that solves these instances in polynomial time.
Furthermore, if we have a {\pnConstant} partial neighborhood system, $\mathcal{C}$ is bounded by a constant, and the functions $\weightsSum$ and $\minWeights$ can be computed in constant time, then the time complexity of such algorithm is $O(|V(G)|)$.
}

\def\COLmainCorollary{
Let $\mathcal{F}$ be a family of graphs of bounded treewidth. Consider a family of instances of a {\rLCproblem{1}} where
\begin{itemize}
\item $G \in \mathcal{F}$,
\item $|\bigcup_{v \in V(G)} L_v|$ is bounded by a constant,
\item $\weightsSum$ and $\minWeights$ can be computed in polynomial time, and
\item $check(v, c)$ can be computed in polynomial time and only depends on $v$, $c(v)$ and the number of neighbors of each color that $v$ has.
\end{itemize}

Then, for such instances, the problem can be solved in polynomial time.
}

\def\THMpolyBoundedTwDeg{
Let $\mathcal{F}$ be a family of graphs of bounded treewidth and bounded degree. Then for any {\rLCproblem{1}} with input graph $G \in \mathcal{F}$, $\mathcal{C}$ polynomial in the input size, and all functions $check$, $\minWeights$ and $\weightsSum$ computable in polynomial time, there exists a polynomial-time algorithm that solves it.
}

\def\LEMpowerBounds{
Let $G$ be a graph and $p \geq 2$. Then
$$\Delta(G) \leq \Delta(G^p) \leq \Delta(G)^p$$
and
$$\max(tw(G), \Delta(G)) \leq tw(G^p) \leq (tw(G)+1)(\Delta(G) + 1)^{\lceil\frac{p}{2}\rceil} - 1.$$
}

\def\COLpolyDistBoundedTwDeg{
Let $\mathcal{F}$ be a family of graphs of bounded treewidth and bounded degree. Then, for any {\rLCproblem{r}} with $G \in \mathcal{F}$, $\mathcal{C}$ polynomial in the input size, and all functions $check$, $\minWeights$ and $\weightsSum$ computable in polynomial time, there exists a polynomial-time algorithm that solves it.
}

\begin{document}
\def\BuenosAires{Buenos Aires, Argentina}
\def\UBA{Universidad de Buenos Aires}
\def\DCFCENUBA{\UBA. Facultad de Ciencias Exactas y Naturales. Departamento de Computaci\'on. \BuenosAires.}
\def\ICC{CONICET-\UBA. Instituto de Investigaci\'on en Ciencias de la Computaci\'on (ICC). \BuenosAires.}

\title[A new approach on locally checkable problems]{A new approach on locally checkable problems}

\author[F. Bonomo-Braberman]{Flavia Bonomo-Braberman}
\address{{\DCFCENUBA} / {\ICC}} \email{fbonomo@dc.uba.ar}

\author[C.L. Gonzalez]{Carolina Luc{\'i}a Gonzalez}
\address{{\ICC}} \email{cgonzalez@dc.uba.ar}

\begin{abstract}
By providing a new framework, we extend previous results on locally checkable problems in bounded treewidth graphs. As a consequence, we show how to solve, in polynomial time for bounded treewidth graphs, double Roman domination and Grundy domination, among other problems for which no such algorithm was previously known.
Moreover, by proving that fixed powers of bounded degree and bounded treewidth graphs are also bounded degree and bounded treewidth graphs, we can enlarge the family of problems that can be solved in polynomial time for these graph classes, including distance coloring problems and distance domination problems (for bounded distances).
\end{abstract}



\maketitle

\section{Introduction}
\label{sec:introduction}

Many combinatorial optimization problems in graphs can be classified as vertex partitioning problems. The partition classes have to verify inner-properties and/or inter-properties, and there is an objective function to minimize or maximize. Some of these properties are \emph{locally checkable}, that is, the property that each vertex has to satisfy with respect to the partition involves only the vertex and its neighbors. This is the case of stable set, dominating set and $k$-coloring, among others.

In the spirit of generalizing this kind of problems,
in~\cite{bodlaender1987,bodlaender1988} Bodlaender defined the \emph{\LCCfull} ({\LCC}) and \emph{\ECCfull} ({\ECC}) problems, and showed polynomial-time algorithms to solve {\LCC} problems on bounded treewidth and bounded degree graphs, and {\ECC} problems on bounded treewidth graphs.
In~\cite{LCVSVP-thesis} Telle defined the \emph{\LCVPfull} ({\LCVP}) problems and in~\cite{LCVSVP-paper2}, Bui-Xuan, Telle and Vatshelle presented dynamic programming algorithms for {\LCVP} problems that run in polynomial time on many graph classes, including interval graphs, permutation graphs and Dilworth $k$ graphs, and in fixed-parameter single-exponential time parameterized by boolean-width.
In~\cite{Cattaneo-Psigma}, Cattan{\'{e}}o and Perdrix defined a different generalization of {\LCVP} problems that allows us to deal with properties of the subset that are not necessarily locally checkable, as for example being connected, and prove hardness results for {\LCVP} problems and such generalizations.

In this paper, we define a new framework for locally checkable problems, which we call \emph{\rLCproblems{r}}.
In a {\rLCproblem{r}}, every vertex $v$ has a list of colors $L_v$ that it can receive, along with the cost $\textsc{w}_{v,i}$ of receiving each color $i \in L_v$. There is a function $check$ that, for each vertex $v$ and each coloring $c$ of the closed neighborhood of radius $r$ of $v$, determines if the colors assigned to $v$ and the vertices at distance at most $r$ from $v$ are permitted for $v$.
We include edge labels $\ell_e$, whose values may be involved in the checking functions.
For technical reasons, other simple operators are also required, such as one that combines the costs and one that compares them.
This approach generalizes {\LCVP} problems, including other problems such as $\{k\}$-domination (which cannot be expressed as a {\LCVP} problem, at least not in an straightforward way).
We further consider a set of global properties (which might be, for example, that certain sets of the partition induce a connected or an acyclic subgraph). In this way, {\ECC} problems can be modeled within our framework.

%--------------------------
A key notion to our paper is the \emph{treewidth} of a graph, which was introduced by Robertson and Seymour~\cite{GraphMinorsIII} (and previously considered under different names by Bertel{\`e} and Brioschi~\cite{B-B-treewidth} and Halin~\cite{Hal-tw}). Graphs of treewidth at most $k$ are called {\em partial $k$-trees}. Some graph classes with bounded treewidth include: forests (treewidth 1); pseudoforests, cacti, outerplanar graphs, and series-parallel graphs (treewidth at most~2); Halin graphs and Apollonian networks (treewidth at most~3)~\cite{GraphMinorsIII,BODLAENDER19981,treewidth-kloks,graph-classes-survey}. In addition, control flow graphs arising in the compilation of structured programs also have bounded treewidth (at most~6)~\cite{THORUP1998159}.

To solve {\rLCproblems{1}} in bounded treewidth graphs, we give an algorithm based on a rather simple computation of a recursive function traversing a special tree decomposition, using dynamic programming as it is usual with this kind of problems, but with an abstraction of the ``extra'' parameters involved in ad-hoc solutions of locally checkable problems. In order to formally describe this abstraction, we introduce the concept of \emph{partial neighborhoods}. A \emph{partial neighborhood system} gives us tools to accumulate information from the neighbors of a vertex. According to the sizes of the sets and the time complexity of the functions involved, we distinguish \emph{polynomial} and \emph{constant} partial neighborhood systems. Then our algorithm is polynomial when some mild conditions are satisfied.
The main result of this paper is the following.

\begin{theoremRep}[\ref{thm:mainTheorem}]
\THMmainTheorem
\end{theoremRep}

Furthermore, by proving that fixed powers of bounded degree and bounded treewidth graphs are also bounded degree and bounded treewidth graphs, we can enlarge the family of problems that can be solved in polynomial time for these graph classes, including distance coloring problems (packing chromatic number~\cite{eccentric-color-trees,broadcast-chromatic,BRESAR20072303,FIALA2010771}, $L(p,1)$-coloring~\cite{L21-1996,L21-1992,CHANG200353,Lp1-2005}), distance independence~\cite{distance-d-indep}, distance domination~\cite{FundamentalsDom}, and distance {\LCVP} problems~\cite{gen-dist-dom-mim-width-2019}, for bounded distances.
These results are unified in the following corollary of Section~\ref{sec:boundedDegree}.

\begin{corollaryRep}[\ref{col:polyDistBoundedTwDeg}]
\COLpolyDistBoundedTwDeg
\end{corollaryRep}

We also prove that NP-complete problems can be reduced to some {\rLCproblems{1}} in complete graphs, even when restricting the sets of colors and edge labels to $\{0,1\}$.
Thus, a generalization of the polynomiality to bounded clique-width graphs is not possible unless P=NP.

We show how to model double Roman domination, minimum chromatic violation and Grundy domination as {\rLCproblems{1}} with polynomial partial neighborhood systems. As a result, we obtain polynomial-time algorithms to solve these problems for bounded treewidth graphs. Until the date and to the best of our knowledge, no such algorithms were previously known.

Courcelle's celebrated theorem (see~\cite{COURCELLE199349}) states that every graph problem definable in Monadic Second-Order (MSO) logic can be solved in linear time for bounded treewidth graphs.
However, its main drawback is that the multiplicative constants in the running time of the algorithm generated with an MSO-formula can be extremely large~\cite{FRICK20043}.
In contrast, the statement of our problem is closer to natural language, and the time complexity of the algorithm for bounded treewidth graphs is fully detailed and involves relatively small constants.

A similar approach for a family of problems of different nature was presented in~\cite{tw-isw}. Namely, the authors define the framework of \emph{algebraic path problems}, enclosing weighted shortest path, dataflow problems, regular expressions, and other problems arising in program analysis in the area of software engineering, and they present an algorithm to solve this framework of problems in concurrent systems such that each of the components is a bounded treewidth graph.

The present paper is organized as follows.
Section~\ref{sec:preliminaries} contains the necessary preliminaries and basic definitions.
The central notion of the paper, {\rLCproblems{r}}, is formally introduced in Section~\ref{sec:problemDescription}.
In Section~\ref{sec:completeGraphs}, we study {\rLCproblems{1}} in complete graphs, under different hypothesis.
In Section~\ref{sec:boundedTreewidth} we give an algorithm to solve {\rLCproblems{1}} parameterized by treewidth.
In Section~\ref{sec:boundedDegree} we analyze the time complexity of {\rLCproblems{r}} in bounded treewidth and bounded degree graphs, and prove that fixed powers of such graphs are also bounded treewidth and bounded degree graphs.
In Section~\ref{sec:globalProp} we extend the algorithm from Section~\ref{sec:boundedTreewidth} with some global properties. In particular, this recovers the results in~\cite{bodlaender1987,bodlaender1988} for {\LCC} problems.
In Section~\ref{sec:examples} we show how to model different problems as {\rLCproblems{1}} with polynomial partial neighborhood systems, obtaining polynomial-time algorithms to solve these problems for bounded treewidth graphs.
We include in Appendix~\ref{sec:problemsDef} the definition of the locally checkable problems mentioned throughout the article.

\section{Basic definitions and preliminary results}
\label{sec:preliminaries}

%---------------------------------------------------------------------------------------------------
% Sets
\subsection{Algebraic definitions}
%---------------------------------------------------------------------------------------------------
% Binary operations
Let $S$ be a set. A \emph{closed binary operation on $S$} is a function $\star \colon S \times S \to S$. It is usual to write $\star(s_1, s_2)$ as $s_1 \star s_2$. Such an operation is \emph{commutative} if $s_1 \star s_2 = s_2 \star s_1$ for all $s_1, s_2 \in S$, and it is \emph{associative} if $(s_1 \star s_2) \star s_3 = s_1 \star (s_2 \star s_3)$ for all $s_1, s_2, s_3 \in S$. An element $e \in S$ is \emph{neutral} (also called \emph{identity}) if $e \star s = s \star e = s$ for all $s \in S$. An element $a \in S$ is \emph{absorbing} if $a \star s = s \star a = a$ for all $s \in S$. It is easy to prove that if $s \in S$ is a neutral (resp. absorbing) element, then this element is unique.
A commutative and associative operation $\star$ can be naturally extended to any nonempty finite subset of $S$, writing $\bigstar_{x \in X} P(x)$ when $\setst{P(x)}{x \in X} \subseteq S$ and $X$ is finite and nonempty, moreover, if the operation also has a neutral element $e$ then we define $\bigstar_{x \in \emptyset} P(x) = e$.

%---------------------------------------------------------------------------------------------------
% Monoids
Let $S$ be a set and $\star$ be a closed binary operation on $S$. Then $(S, \star)$ is a \emph{monoid} if $\star$ is associative and has a neutral element. If $\star$ is also commutative then $(S, \star)$ is a \emph{commutative monoid}.

%---------------------------------------------------------------------------------------------------
% Binary relations and partially/totally ordered sets
A \emph{binary relation $\mathcal{R}$ on a set $S$} is a subset of the Cartesian product $S \times S$. It is usual to write $(s_1, s_2) \in \mathcal{R}$ as $s_1 \mathcal{R} s_2$. We say that $\mathcal{R}$ is \emph{reflexive} if $s \mathcal{R} s$ for all $s \in S$, \emph{antisymmetric} if $s_1 \mathcal{R} s_2 \land s_2 \mathcal{R} s_1 \Rightarrow s_1 = s_2$ for all $s_1, s_2 \in S$, and \emph{transitive} if $s_1 \mathcal{R} s_2 \land s_2 \mathcal{R} s_3 \Rightarrow s_1 \mathcal{R} s_3$ for all $s_1, s_2, s_3 \in S$. If $\mathcal{R}$ is reflexive, antisymmetric and transitive, then $(S, \mathcal{R})$ is called a \emph{partial order} (or \emph{partially ordered set}). If in addition $s_1 \mathcal{R} s_2 \lor s_2 \mathcal{R} s_1$ for every $s_1, s_2 \in S$, then $(S, \mathcal{R})$ is a \emph{total order} (or \emph{totally ordered set}).

%---------------------------------------------------------------------------------------------------
% Minimum operation and maximum element
Let $(S, \preceq)$ be a totally ordered set. A \emph{maximum element} is an element $m \in S$ such that $s \preceq m$ for all $s \in S$. Note that not every totally ordered set has a maximum element, and it is easy to prove that if it does have a maximum element then this element is unique. The \emph{minimum operation}, $\min$, is the closed binary operation on $S$ such that $\min(s_1, s_2) = s_1$ if $s_1 \preceq s_2$ and $\min(s_1, s_2) = s_2$ if $s_2 \preceq s_1$. It is easy to prove that $\min$ is commutative and associative.

%---------------------------------------------------------------------------------------------------
% Finite and co-finite sets
A set of natural numbers is \emph{co-finite} if its complement with respect to the set of natural numbers is finite.

%---------------------------------------------------------------------------------------------------
% Notation [a,b]
We denote by $\intInterval{a}{b}$, with $a,b \in \mathbb{Z}$ and $a < b$, the set of all integer numbers greater than or equal to $a$ and less than or equal to $b$, that is $\{a, a+1, \ldots, b\}$.

%---------------------------------------------------------------------------------------------------
% Weight of a function
Given a set $S$ and a function $f \colon S \to \mathbb{R}$, the \emph{weight} of the function $f$ (finite or infinite) is defined as $f(S) = \sum_{s \in S} f(s)$.

%---------------------------------------------------------------------------------------------------
% Bool
Throughout this paper we will work with the set $\BoolSet = \{\True, \False\}$ of boolean values and all the usual logical operators, such as $\neg$, $\land$, $\lor$ and $\Rightarrow$.

%---------------------------------------------------------------------------------------------------
% Function restriction
Let $f \colon A \rightarrow B$ be a function and let $S \subseteq A$. We denote by $f|_S$ the function $f$ restricted to the domain $S$, that is, the function $f|_S \colon S \to B$ is defined as $f|_S(s) = f(s)$ for all $s \in S$.

%---------------------------------------------------------------------------------------------------
% Finite-State Automata
\subsection{Automata}
A \emph{deterministic finite-state automaton} is a five-tuple $(Q, \Sigma, \delta, q_0, F)$ that consists of
\begin{itemize}
\item $Q$: a finite set of \emph{states},
\item $\Sigma$: a finite set of \emph{input symbols} (often called the \emph{alphabet}),
\item $\delta \colon Q\times \Sigma \rightarrow Q$: a \emph{transition function},
\item $q_0 \in Q$: an \emph{initial} or \emph{start state}, and
\item $F \subseteq Q$: a set of \emph{final} or \emph{accepting states}.
\end{itemize}

We say that an automaton $M = (Q, \Sigma, \delta, q_0, F)$ \emph{accepts} a string $s_1{\ldots}s_n$, with $n \geq 1$, if and only if $s_i \in \Sigma$ for all $1 \leq i \leq n$ and $\delta(\ldots \delta(\delta(q_0, s_1), s_2) \ldots , s_n) \in F$.

For example, the automaton $M = (\{q_0, q_1\}, \{1\}, \delta, q_0, \{q_1\})$ where $\delta(q_0, 1) = q_1$ and $\delta(q_1, 1) = q_0$, is an automaton that accepts sequences of an odd number of $1$s.

For more about automata theory we refer the reader to \cite{HopcroftUllman}.

%---------------------------------------------------------------------------------------------------
% Computability
\subsection{Computability}
A function is \emph{polynomial time computable} if there exists an algorithm $A$ and a polynomial $p(n)$ such that, for all inputs of length $n$, $A$ computes the function and runs in time less than or equal to $p(n)$.

All sets considered in this article contain elements that can be encoded and passed as parameters to a computable function.

We will not delve further into this topic. For more information we refer the reader to the vast literature on computability theory.

%---------------------------------------------------------------------------------------------------
% Graphs: basic definitions
\subsection{Basic definitions on graphs}
%---------------------------------------------------------------------------------------------------
% Graph, neighbors, degree
Let $G$ be a finite, simple and undirected graph. We denote by $V(G)$ and $E(G)$ the vertex set and edge set, respectively, of $G$.
For any $W\subseteq V(G)$, we denote by $G[W]$ the subgraph of $G$ induced by $W$. Let $N_G(v)$ (\emph{open neighborhood of $v$}) be the set of neighbors of $v \in V(G)$ and let $N_G[v] = N_G(v) \cup \{v\}$ (\emph{closed neighborhood of $v$}).
The \emph{closed neighborhood of a set $S$} is $N_G[S] = \bigcup_{v \in S}N_G[v]$.
The \emph{degree} of a vertex $v$ is $d_G(v) = |N_G(v)|$. The maximum degree of a vertex in $G$ is denoted by $\Delta(G)$.

%---------------------------------------------------------------------------------------------------
% Graph class
A \emph{graph class} is a collection of graphs that is closed under isomorphism. Given a graph class $\mathcal{G}$, we say that $\mathcal{G}$ is \emph{of bounded degree} if $\sup\{ \Delta(G) \mid G \in \mathcal{G} \} < \infty$.

%---------------------------------------------------------------------------------------------------
% Connected graph, connected component, distance, power
A graph $G$ is \emph{connected} if for every pair of vertices $u, v$ in $V(G)$ there exists a path in $G$ from $u$ to $v$.
A \emph{connected component} of a graph is an inclusion-wise maximal connected subgraph of it.
For two vertices $x,y$ in a connected graph $G$, we denote by $\mbox{dist}_G(x,y)$ the \emph{distance between $x$ and $y$}, that is, the length (number of edges) of a shortest $x,y$-path in $G$.
Let $N_G^k[v]$ be the set of vertices at distance at most $k$ from $v$ in $G$, and $N_G^k(v) = N_G^k[v] - \{v\}$.
Let $M_G^k(v)$ be the set of edges whose endpoints at distance at most $k$ from $v$ in $G$.
The \emph{$k$-th power of $G$} is the graph denoted by $G^k$ such that for all distinct vertices $x,y$ in $V(G)$, $x$ is adjacent to $y$ in $G^k$ if and only if $\mbox{dist}_G(x,y) \leq k$.

%---------------------------------------------------------------------------------------------------
% Complete graph, clique, stable set
A \emph{complete graph} is a graph whose vertices are pairwise adjacent. We denote by $K_r$ the complete graph on $r$ vertices. A \emph{clique} (resp. \emph{stable set} or \emph{independent set}) in a graph is a set of pairwise adjacent (resp. nonadjacent) vertices. The maximum size of a clique (resp. independent set) in the graph $G$ is denoted by $\omega(G)$ (resp. $\alpha(G)$).

%---------------------------------------------------------------------------------------------------
% Bipartite graph, complete bipartite, star
A graph $G$ is \emph{bipartite} if $V(G)$ can be partitioned into two stable sets $V_1$ and $V_2$, and $G$ is \emph{complete bipartite} if every vertex of $V_1$ is adjacent to every vertex of $V_2$. We denote by $K_{r,s}$ the complete bipartite graph with $|V_1|=r$ and $|V_2|=s$. The \emph{star} $S_n$ is the complete bipartite graph $K_{1,n-1}$.

%---------------------------------------------------------------------------------------------------
% Coloring, chromatic number, list coloring
A \emph{proper $k$-coloring} of a graph is a partition of its vertices into at most $k$ stable sets, each of them called \emph{color class}. Equivalently, a proper $k$-coloring is an assignment of colors to vertices such that adjacent vertices receive different colors, and the number of colors used is at most $k$. The \emph{chromatic number} $\chi(G)$ of a graph $G$ is the minimum $k$ that allows a proper $k$-coloring of $G$. In the more general \textsc{List-coloring} problem, each vertex $v$ has a list $L(v)$ of available colors for it.

%---------------------------------------------------------------------------------------------------
% Domination
A pair of vertices or a pair of edges \emph{dominate} each other when they are either equal or adjacent, while a vertex and an edge \emph{dominate} each other when the vertex belongs to the edge. We will denote by $\gamma_{U,W}(G)$, for $U, W$ sets of elements of $G$, the minimum cardinality or weight of a subset $S$ of $U$ which \emph{dominates} $W$. The parameter $\gamma_{V,V}$ is also denoted simply by $\gamma$, and the associated problem is known as \textsc{Minimum Dominating Set}. Parameters $\gamma_{V,E}$ and $\gamma_{E,V}$ are associated with the \textsc{Minimum Vertex Cover} and \textsc{Minimum Edge Cover}, respectively.
In the \textsc{Minimum $\{k\}$-domination} problem, given a graph $G$ we want to find the minimum weight of a function $f \colon V(G) \to \{0, 1, \ldots, k\}$ such that $\sum_{u \in N_G[v]} f(u) \geq k$ for all $v \in V(G)$.

%---------------------------------------------------------------------------------------------------
% Subdivision graph
For a graph $G$ and $uv \in E(G)$, the graph obtained by \emph{subdividing $uv$ in $G$} arises from $G$ by adding a new vertex $w$, making $w$ adjacent to $u$ and $v$, and then deleting the edge $uv$.
The \emph{subdivision graph} of $G$, obtained by subdividing each of the edges of $G$, is $S(G) = (V', E')$ where $V' = V(G) \cup E(G)$ and $E' = \{ve : v \in V(G), e \in E(G)\text{, and } \text{$v$ is an endpoint of $e$}\}$.

%---------------------------------------------------------------------------------------------------
% Jagged graph
The \emph{jagged graph} of $G$ is $J(G) = (V', E')$ where $V' = V(G) \cup E(G)$ and $E' = E(G) \cup \{ve : v \in V(G), e \in E(G)\text{, and } \text{$v$ is an endpoint of $e$}\}$.

%---------------------------------------------------------------------------------------------------
% Line graph
The \emph{line graph} of a graph $G$ is denoted by $L(G)$ and has as vertex set $E(G)$, where two vertices are adjacent in $L(G)$ if and only if the corresponding edges have a common endpoint (i.e., are adjacent) in $G$. The \emph{total graph} of $G$, denoted by $T(G)$, is defined similarly: its vertex set is $V(G) \cup E(G)$, $V(G)$ induces $G$, $E(G)$ induces $L(G)$, and $v \in V(G)$, $uw \in E(G)$ are adjacent in $T(G)$ if and only if either $v = u$ or $v = w$.

%---------------------------------------------------------------------------------------------------
% Acyclic graph, forest, tree, root, parent, child, binary tree
A graph, or a subgraph of a graph, is \emph{acyclic} if it does not contain a cycle of length at least three. An acyclic graph is called a \emph{forest}. A \emph{tree} is a connected acyclic graph. In a tree $T$, we usually call the elements in $V(T)$ \emph{nodes}. A node with degree at most~1 is called a \emph{leaf} and a node of degree at least~2 is called an \emph{internal node}. A tree is called a \emph{rooted tree} if one vertex has been designated the \emph{root}, in which case the edges have a natural orientation, towards or away from the root. For a rooted tree $T$ and $u \in V(T)$, the neighbor of $u$ on the path to the root is called the \emph{parent} of $u$ and a vertex $v$ is a \emph{child} of $u$ if $u$ is the parent of $v$. A \emph{binary tree} is a rooted tree where every internal node has at most two children.

%---------------------------------------------------------------------------------------------------
% Treewidth
\subsection{Definitions and preliminary results on treewidth}
A \emph{tree-decomposition} of a graph $G$ is a family $\setst{X_i}{i \in I}$ of subsets of $V(G)$ (called \emph{bags}), together with a tree $T$ with $V(T) = I$, satisfying the following properties:
\begin{itemize}
\item[(W1)] $\bigcup_{i \in I} X_i = V(G)$.
\item[(W2)] Every edge of $G$ has both its ends in $X_i$ for some $i \in I$.
\item[(W3)] For all $v \in V(G)$, the set of nodes $\setst{i \in I}{v \in X_i}$ induces a subtree of $T$.
\end{itemize}

The \emph{width} of the tree-decomposition is $\max\setst{|X_i| - 1}{i \in I}$. The \emph{treewidth} of $G$, denoted $tw(G)$, is the minimum $w \geq 0$ such that $G$ has a tree-decomposition of width less or equal $w$.

Given a graph class ${\mathcal G}$, the treewidth of $\mathcal{G}$ is $tw(\mathcal{G}) = \sup\{tw(G) \mid G \in \mathcal{G}\}$. We say that $\mathcal{G}$ is \emph{of bounded treewidth} if $tw(\mathcal{G}) < \infty$.

We will often make use of the following basic properties of the treewidth, some of which can be easily deduced.

\begin{proposition}\label{prop:twedges}
Let $\mathcal{G}$ be a family of graphs of bounded treewidth. If $G \in \mathcal{G}$ then $|E(G)|$ is $O(|V(G)|)$.
\end{proposition}

\begin{proposition}\label{prop:twsubgraph}
If $H$ is a subgraph of a graph $G$ then $tw(H)\leq tw(G)$.
\end{proposition}

\begin{theorem}[{\cite[pages 1 and 2]{GraphMinorsII}}]\label{thm:twclique}
For every graph $G$, $tw(G) \geq \omega(G)-1$. Moreover, $tw(G) \geq \chi(G)-1$.
\end{theorem}

\begin{theorem}[{\cite[page 76]{treewidth-kloks}}]\label{thm:twsubdiv}
The treewidth of $S(G)$ is equal to the treewidth of $G$.
\end{theorem}

\begin{theorem}\label{thm:twjagged}
The treewidth of $J(G)$ is less than or equal to $tw(G)+1$.
\end{theorem}

A tree-decomposition $(T, \{X_t\}_{t \in V(T)})$ is \emph{nice}~\cite[Definition 13.1.4]{treewidth-kloks} if
\begin{itemize}
\item $T$ is a rooted binary tree;
\item if a node $i$ has two children $j$ and $k$ then $X_i = X_j = X_k$;\hfill{(\emph{join} node)}
\item if a node $i$ has one child $j$, then either
    \begin{itemize}
    \item $|X_i| = |X_j| - 1$ and $X_i \subset X_j$, or\hfill{(\emph{forget} node)}
    \item $|X_i| = |X_j| + 1$ and $X_i \supset X_j$.\hfill{(\emph{introduce} node)}
    \end{itemize}
\end{itemize}

Let $T_i$ be the subtree of $T$ rooted at node $i$. We will denote by $G_i$ the subgraph of $G$ induced by $\bigcup_{j \in V(T_i)}X_j$.

\begin{theorem}[{\cite[Theorem 1]{5aproxTW}}]\label{thm:treedec}
There exists an algorithm, that given an $n$-vertex graph $G$ and an integer $k$, in time $O(c^k n)$ for some $c \in \mathbb{N}$, either outputs that the treewidth of $G$ is larger than $k$, or constructs a tree-decomposition of $G$ of width at most $5k + 4$.
\end{theorem}

\begin{theorem}[{\cite[Lemma 13.1.3]{treewidth-kloks}}]\label{thm:nicetreedec}
For constant $k$, given a tree-decomposition of a graph $G$ of width $k$ and $O(n)$ nodes, where $n$ is the number of vertices of $G$, one can find a nice tree-decomposition of $G$ of width $k$ and with at most $4n$ nodes in $O(n)$ time.
\end{theorem}

However, we will work with a slight modification of nice tree decompositions, where the bags of the root and leaves have only one vertex each.
\begin{definition}\label{def:easytreedec}
A tree-decomposition $(T, \{X_t\}_{t \in V(T)})$ is called \emph{easy} if
\begin{itemize}
\item $T$ is a binary tree rooted at $r$ such that $|X_r| = 1$;
\item if a node $i$ has two children $j$ and $k$ then $X_i = X_j = X_k$;\hfill{(\emph{join} node)}
\item if a node $i$ has one child $j$, then either
    \begin{itemize}
    \item $|X_i| = |X_j| - 1$ and $X_i \subset X_j$, or\hfill{(\emph{forget} node)}
    \item $|X_i| = |X_j| + 1$ and $X_i \supset X_j$;\hfill{(\emph{introduce} node)}
    \end{itemize}
\item if a node $i$ has no children, then $|X_i| = 1$.\hfill{(\emph{leaf} node)}
\end{itemize}
\end{definition}

It is straightforward to prove that, given a nice tree-decomposition $(T, \{X_t\}_{t \in V(T)})$ of width $k$ and $O(n)$ nodes, one can construct in $O(kn)$ time an easy tree-decomposition of width $k$ and $O(kn)$ nodes.

%---------------------------------------------------------------------------------------------------
% Locally checkable problems
\subsection{Definitions and preliminary results on frameworks for locally checkable problems}
Throughout this article we will make special emphasis on two previous frameworks for locally checkable problems.
In this section we review their definitions and results related to our work.

%---------------------------------------------------------------------------------------------------
% LC-VSVP problems
\subsubsection{{\LCVP} problems}
Let $\sigma$ and $\rho$ be finite or co-finite subsets of non-negative integer numbers. A subset $S$ of vertices of a graph $G$ is a \emph{sigma-rho set}, or simply $(\sigma,\rho)$-set, of $G$ if for every $v$ in $S$, $|N(v) \cap S| \in \sigma$, and for every $v$ in $V(G)\setminus S$, $|N(v) \cap S| \in \rho$. The \emph{locally checkable vertex subset problems}~\cite{LCVSVP-thesis} consist of finding a minimum or maximum $(\sigma,\rho)$-set in an input graph $G$, possibly on vertex weighted graphs.

A generalization of these problems asks for a partition of $V(G)$ into $q$ classes, with each class satisfying a certain $(\sigma, \rho)$-property, as follows. A \emph{degree constraint matrix} $D_q$ is a $q \times q$ matrix with entries being finite or co-finite subsets of non-negative integer numbers. A $D_q$-partition of a graph $G$ is a partition $\{V_1, V_2, \dots, V_q\}$ of $V(G)$ such that for $1 \leq i, j \leq q$ it holds that for every $v \in V_i$, $|N(v) \cap V_j| \in D_q[i, j]$. A \emph{{\LCVPfull} ({\LCVP}) problem}~\cite{LCVSVP-thesis} consists of deciding if $G$ has a $D_q$ partition. Optimization versions can be defined, possibly on vertex weighted graphs.

The \emph{distance-$r$ {\LCVPfull} problems}~\cite{gen-dist-dom-mim-width-2019} further generalize {\LCVP} problems by considering, for each vertex $v$, $N_G^r(v)$ instead of $N_G(v)$.

In~\cite{LCVSVP-paper2}, Bui-Xuan, Telle and Vatshelle presented dynamic programming algorithms for {\LCVP} problems that run in polynomial time on many graph classes, including interval graphs, permutation graphs and Dilworth $k$ graphs, and in fixed-parameter single-exponential time parameterized by boolean-width.
In~\cite{gen-dist-dom-mim-width-2019} Jaffke, Kwon, Str{\o}mme and Telle presented dynamic programming algorithms for distance-$r$ {\LCVPfull} problems in graphs of bounded mim-width.

%---------------------------------------------------------------------------------------------------
% LCC-ECC problems
\subsubsection{{\LCC} and {\ECC} problems}
In~\cite{bodlaender1987,bodlaender1988} Bodlaender defined the \emph{\LCCfull} ({\LCC}) and \emph{\ECCfull} ({\ECC}) problems, and showed polynomial-time algorithms to solve {\LCC} problems on bounded treewidth and bounded degree graphs, and {\ECC} problems on bounded treewidth graphs.

%---------------------------------------------------------------------------------------------------
\begin{definition}[{\cite[Definition 2.9]{bodlaender1987}}]
Let $\Pi$ be a graph decision problem, and let $D_{\Pi}$ be the set of instances of $\Pi$, $Y_{\Pi}$ the set of instances for which the answer is ``yes'', and $s : D_{\Pi} \to \mathbb{N}$ be a function that assigns sizes to instances of $\Pi$. We say $\Pi$ is a \emph{basic local condition problem}, if and only if there exist
\begin{itemize}
\item non-negative integers $m, c \in \mathbb{N}$,
\item $m$ commutative monoids $(M^1, \oplus^1), \ldots, (M^m, \oplus^m)$, and
\item a tuple $(M^{m+1}, \oplus^{m+1}, \preceq)$ such that
$(M^{m+1}, \oplus^{m+1})$ is a commutative monoid,
$(M^{m+1}, \preceq)$ is a total order
and $a \preceq b \Rightarrow a \oplus^{m+1} c \preceq b \oplus^{m+1} c$ for all $a,b,c \in M^{m+1}$
\end{itemize}
such that
\begin{itemize}
\item each $D \in D_{\Pi}$ is of the form $(G, (X, Y, R_1, \ldots, R_m, K, I))$, where
	\begin{itemize}
	\item $G$ is an undirected graph
	\item $X$ is a finite set with $s(D) \geq |X|$
	\item $Y$ is a finite set with $s(D) \geq |Y|$
	\item for all $i$, $1 \leq i \leq m$, $R_i$ denotes a subset of $M^i$
	\item $K \in M^{m+1}$
	\end{itemize}

\item for all $i$, $1 \leq i \leq m+1$, there exists a function $val_i$, that maps all 4-tuples, consisting of an instance $D$ $=$ $(G,$ $(X,$ $Y,$ $R_1,$ $\ldots,$ $R_m,$ $K,$ $I))$ $\in D_{\Pi}$, a vertex $v \in V(G)$, and functions $f \colon N_G^c[v] \rightarrow X$, $g \colon M_G^c(v) \rightarrow Y$, to elements of $M_i$, such that for all (constants) $d \in \mathbb{N}^{+}$:
	\begin{enumerate}
	\item there exists an algorithm that calculates $val_i(D, v, f, g)$, for all $D$ $=$ $(G,$ $(X,$ $Y,$ $R_1,$ $\ldots,$ $R_m,$ $K,$ $I))$ $\in D_{\Pi}$, $v \in V(G)$, $f \colon N_G^c[v] \rightarrow X$, $g \colon M_G^c(v) \rightarrow Y$ with $\text{degree}(G) \leq d$, in time, polynomial in $s(D)$.
	\item if $1 \leq i \leq m$, there is a polynomial $p_i$, such that for all $D$ $=$ $(G,$ $(X,$ $Y,$ $R_1,$ $\ldots,$ $R_m,$ $K,$ $I))$ $\in D_{\Pi}$, with $\text{degree}(G) \leq d$ and subsets $W \subseteq V(G) : |\{\bigoplus^i_{w \in W} val_i(D, w, f|_{N_G^c[w]}, g|_{M_G^c(w)}) \;|\; f \colon N_G^c[W] \rightarrow X, g \colon M_G^c(W) \rightarrow Y\}| \leq p_i(s(D))$.
	\item there exists an algorithm that calculates $a \oplus^i b$ for given $a, b$, such that there are $D$ $=$ $(G,$ $(X,$ $Y,$ $R_1,$ $\ldots,$ $R_m,$ $K,$ $I))$ $\in D_{\Pi}$, with $\text{degree}(G) \leq d$, $W_1 \subseteq V(G)$, $W_2 \subseteq V(G)$, $W_1 \cap W_2 = \emptyset$, $f \colon N_G^c[W_1 \cup W_2] \rightarrow X$, $g \colon M_G^c(W_1 \cup W_2) \rightarrow Y$, $a = \bigoplus^i_{w \in W_1} val_i(D, w, f|_{N_G^c[w]}, g|_{M_G^c(w)})$ and $b = \bigoplus^i_{w \in W_2} val_i(D, w, f|_{N_G^c[w]}, g|_{M_G^c(w)})$, in time, polynomial in $s(D)$.
	\item if $i = m+1$, then there exists an algorithm, that calculates whether $a \preceq b$ for given $a,b$, such that there are $D$ $=$ $(G,$ $(X,$ $Y,$ $R_1,$ $\ldots,$ $R_m,$ $K,$ $I))$ $\in D_{\Pi}$, with $\text{degree}(G) \leq d$, $W \subseteq V(G)$, $f_1 \colon N_G^c[W] \rightarrow X$, $f_2 \colon N_G^c[W] \rightarrow X$, $g_1 \colon M_G^c(W) \rightarrow Y$, $g_2 \colon M_G^c(W) \rightarrow Y$, $a = \bigoplus^{m+1}_{w \in W} val_{m+1}(D, w, f_1|_{N_G^c[w]},$ $g_1|_{M_G^c(w)})$ and $b = \bigoplus^{m+1}_{w \in W} val_{m+1}(D, w, f_2|_{N_G^c[w]}, g_2|_{M_G^c(w)})$ or $b = K$, in time polynomial in $s(D)$.
	\item if $1 \leq i \leq m$, there exists an algorithm that calculates for all $D$ $=$ $(G,$ $(X,$ $Y,$ $R_1,$ $\ldots,$ $R_m,$ $K,$ $I))$ $\in D_{\Pi}$ with $\text{degree}(G) \leq d$, $f \colon V(G) \rightarrow X$, $g \colon E(G) \rightarrow Y$ and given $a = \bigoplus^i_{w \in V(G)} val_i(D, w, f|_{N_G^c[w]}, g|_{M_G^c(w)})$ whether $a \in R_i$, in time polynomial in $s(D)$.
	\end{enumerate}

\item For all $D = (G, (X, Y, R_1, \ldots, R_m, K, I)) \in D_{\Pi} : D \in Y_{\Pi}$, if and only if there exists functions $f \colon V(G) \rightarrow X$, $g \colon E(G) \rightarrow Y$, with
	\begin{enumerate}
	\item $\forall i, 1 \leq i \leq m : \bigoplus^i_{v \in V(G)} val_i(D, v, f|_{N_G^c[v]}, g|_{M_G^c(v)}) \in R_i$
	\item $\bigoplus^{m+1}_{v \in V(G)} val_{m+1}(D, v, f|_{N_G^c[v]}, g|_{M_G^c(v)}) \leq K$.
	\end{enumerate}
\end{itemize}
\end{definition}

\begin{definition}[{\cite[Definition 2.10]{bodlaender1987}}]
Let $\Pi$ be a graph decision problem. We say $\Pi$ is a local condition composition problem, if and only if there exists a basic local condition composition problem $\Pi'$ and a graph-invariant polynomial transformation from $\Pi$ to $\Pi'$. The class of local condition composition problems is denoted by {\LCC}.
\end{definition}

%---------------------------------------------------------------------------------------------------
\begin{definition}[{\cite[Definition 2.11]{bodlaender1987}}]
Let $\Pi$ be a graph decision problem, and let $D_{\Pi}$ be the set of instances of $\Pi$, $Y_{\Pi}$ the set of instances for which the answer is ``yes'', and $s : D_{\Pi} \to \mathbb{N}$ be a function that assigns sizes to instances of $\Pi$. We say $\Pi$ is a \emph{basic edge condition problem}, if and only if there exist
\begin{itemize}
\item a non-negative integer $m \in \mathbb{N}$,
\item $m$ commutative monoids $(M^1, \oplus^1), \ldots, (M^m, \oplus^m)$, and
\item a tuple $(M^{m+1}, \oplus^{m+1}, \preceq)$ such that
$(M^{m+1}, \oplus^{m+1})$ is a commutative monoid,
$(M^{m+1}, \preceq)$ is a total order
and $a \preceq b \Rightarrow a \oplus^{m+1} c \preceq b \oplus^{m+1} c$ for all $a,b,c \in M^{m+1}$
\end{itemize}
such that
\begin{itemize}
\item each $D \in D_{\Pi}$ is of the form $(G, (X, Y, R_1, \ldots, R_m, K, I))$, where
	\begin{itemize}
	\item $G$ is an undirected graph
	\item $X$ is a finite set with $s(D) \geq |X|$
	\item $Y$ is a finite set with $s(D) \geq |Y|$
	\item for all $i$, $1 \leq i \leq m$, $R_i$ denotes a subset of $M^i$
	\item $K \in M^{m+1}$
	\end{itemize}

\item for all $i$, $1 \leq i \leq m+1$, there exists a function $val_i$, that maps all 4-tuples, consisting of an instance $D$ $=$ $(G,$ $(X,$ $Y,$ $R_1,$ $\ldots,$ $R_m,$ $K,$ $I))$ $\in D_{\Pi}$, an edge $uv \in E(G)$, and functions $f \colon \{u, v\} \rightarrow X$, $g \colon \{uv\} \rightarrow Y$, to elements of $M_i$, such that:
	\begin{enumerate}
	\item there exists an algorithm that calculates $val_i(D, e, f, g)$, for all $D$ $=$ $(G,$ $(X,$ $Y,$ $R_1,$ $\ldots,$ $R_m,$ $K,$ $I))$ $\in D_{\Pi}$, $uv \in E(G)$, $f \colon \{u,v\} \rightarrow X$, $g \colon \{uv\} \rightarrow Y$ in time, polynomial in $s(D)$.
	\item if $1 \leq i \leq m$, there is a polynomial $p_i$, such that for all $D$ $=$ $(G,$ $(X,$ $Y,$ $R_1,$ $\ldots,$ $R_m,$ $K,$ $I))$ $\in D_{\Pi}$, and subsets $E' \subseteq E(G) : |\{\bigoplus^i_{uv \in E'} val_i(D,$ $uv,$ $f(u),$ $f(v),$ $g(uv)) \;|\; f \colon N_G(E) \rightarrow X, g \colon E \rightarrow Y\}| \leq p_i(s(D))$.
	\item there exists an algorithm that calculates $a \oplus^i b$ for given $a, b$, such that there are $D$ $=$ $(G,$ $(X,$ $Y,$ $R_1,$ $\ldots,$ $R_m,$ $K,$ $I))$ $\in D_{\Pi}$, $E_1 \subseteq E(G)$, $E_2 \subseteq E(G)$, $E_1 \cap E_2 = \emptyset$, $f \colon N_G(E_1 \cup E_2) \rightarrow X$, $g \colon E_1 \cup E_2 \rightarrow Y$, $a = \bigoplus^i_{uv \in E_1} val_i(D,$ $uv,$ $f(u),$ $f(v),$ $g(uv))$ and $b = \bigoplus^i_{uv \in E_2} val_i(D,$ $uv,$ $f(u),$ $f(v),$ $g(uv))$, in time, polynomial in $s(D)$.
	\item if $i = m+1$, then there exists an algorithm, that calculates whether $a \preceq b$ for given $a,b$, such that there are $D$ $=$ $(G,$ $(X,$ $Y,$ $R_1,$ $\ldots,$ $R_m,$ $K,$ $I))$ $\in D_{\Pi}$, $E' \subseteq E(G)$, $f_1 \colon N_G(E') \rightarrow X$, $f_2 \colon N_G(E') \rightarrow X$, $g_1 \colon E' \rightarrow Y$, $g_2 \colon E' \rightarrow Y$, $a = \bigoplus^{m+1}_{uv \in E'} val_{m+1}(D, uv, f_1(u), f_1(v), g_1(uv))$ and $b = \bigoplus^{m+1}_{uv \in E'} val_{m+1}(D, uv, f_2(u), f_2(v), g_2(uv))$ or $b = K$, in time polynomial in $s(D)$.
	\item if $1 \leq i \leq m$, there exists an algorithm that calculates for all $D$ $=$ $(G,$ $(X,$ $Y,$ $R_1,$ $\ldots,$ $R_m,$ $K,$ $I))$ $\in D_{\Pi}$, $f \colon V(G) \rightarrow X$, $g \colon E(G) \rightarrow Y$ and given $a = \bigoplus^i_{uv \in E(G)} val_i(D,$ $uv,$ $f(u),$ $f(v),$ $g(uv))$ whether $a \in R_i$, in time polynomial in $s(D)$.
	\end{enumerate}

\item For all $D = (G, (X, Y, R_1, \ldots, R_m, K, I)) \in D_{\Pi} : D \in Y_{\Pi}$, if and only if there exists functions $f \colon V(G) \rightarrow X$, $g \colon E(G) \rightarrow Y$, with
	\begin{enumerate}
	\item $\forall i, 1 \leq i \leq m : \bigoplus^i_{uv \in E(G)} val_i(D,$ $uv,$ $f(u),$ $f(v),$ $g(uv)) \in R_i$
	\item $\bigoplus^{m+1}_{uv \in E(G)} val_{m+1}(D, uv, f(u), f(v), g(uv)) \leq K$.
	\end{enumerate}
\end{itemize}
\end{definition}

\begin{definition}[{\cite[Definition 2.12]{bodlaender1987}}]
Let $\Pi$ be a graph decision problem. We say $\Pi$ is an edge condition composition problem, if and only if there exists a basic edge condition composition problem $\Pi'$ and a graph-invariant polynomial transformation from $\Pi$ to $\Pi'$. The class of edge condition composition problems is denoted by {\ECC}.
\end{definition}

\begin{theorem}[{\cite[Theorem 2.5]{bodlaender1987}}]
$\ECC \subseteq \LCC$.
\end{theorem}

%---------------------------------------------------------------------------------------------------
\begin{theorem}[{\cite[Theorem 3.7]{bodlaender1987}}]
(i) Let $\Pi \in \LCC$, and let $k, d \in \mathbb{N}^{+}$. Let $\Theta$ be a class of graphs with $G \in \Theta \Rightarrow \text{degree}(G) \leq d \land \text{treewidth}(G) \leq k$. Then $\Pi$ restricted to $\Theta$ can be solved in polynomial time.

(ii) Let $\Pi \in \ECC$, and let $k \in \mathbb{N}^{+}$. Let $\Theta$ be a class of graphs with $G \in \Theta \Rightarrow \text{treewidth}(G) \leq k$. Then $\Pi$ restricted to $\Theta$ can be solved in polynomial time.
\end{theorem}

\section{\rLCproblems{r}}
\label{sec:problemDescription}

Let $r \in \mathbb{N}$.
Let $G$ be a simple undirected graph.
Then suppose we have the following:
%----------------
\begin{itemize}
% Edge labels
\item a set $\LabelSet$ and, for each edge $e \in E(G)$, a label $\ell_{e} \in \LabelSet$;

% Colors
\item a set $\ColorSet$ and, for every vertex $v \in V(G)$, a nonempty set (also called list) $L_{v} \subseteq \ColorSet$ of possible colors for $v$;

% Weights
\item a totally ordered set $(\WeightSet, \wlesseq)$ with a maximum element (called $\Error$), together with the minimum operation of the order $\wlesseq$ (called $\minWeights$) and a closed binary operation on $\WeightSet$ (called $\weightsSum$) that is commutative and associative, has a neutral element (called $\weightsNeutral$) and an absorbing element that is equal to $\Error$, and is such that $s_1 \wlesseq s_2 \Rightarrow s_1 \weightsSum s_3 \wlesseq s_2 \weightsSum s_3$ for all $s_1, s_2, s_3 \in \WeightSet$;

\item for every vertex $v \in V(G)$ and for every color $i \in L_{v}$, a weight (or cost) $\textsc{w}_{v,i} \in \WeightSet - \{\Error\}$ of assigning color $i$ to vertex $v$; and

% Local check function
\item a function $check$ that, given a vertex $v \in V(G)$ and given a color assignment $c \colon N_G^r[v] \rightarrow \bigcup_{u \in N_G^r[v]} L_u$ such that $c(u) \in L_u$ for all $u \in N_G^r[v]$, returns $\True$ if the vertex $v$ together with its neighborhood of radius $r$ (considering the labels of the edges $uv$ with $u \in N_G^r(v)$) satisfies a certain condition, and $\False$ otherwise.
\end{itemize}
%----------------

We say that an assignment of colors to vertices $c$ is \emph{valid in $V$} if $V$ is the domain of $c$ and $c(v) \in L_v$ for all $v \in V$, and it is \emph{proper} if it is valid in $V(G)$ and $check\left(v, c|_{N_G^r[v]}\right)$ is true for every $v \in V(G)$.
The weight of a color assignment $c$ valid in $V$ is $\textsc{w}(c) = \bigweightsSum_{v \in V} \textsc{w}_{v, c(v)}$.

Given all the previously defined $G$, $\ell_{e}$, $L_v$, $(\WeightSet, \wlesseq, \weightsSum)$, $\textsc{w}_{v,i}$ and $check$, an \emph{\rLCproblem{r}} consists of finding the minimum weight (according to the order $\wlesseq$) of a proper assignment of colors to vertices.
If no such coloring exists, the answer should be $\Error$.

If we further consider a set $\globalPropSet(c)$ of global properties (such as ``the subgraph induced by the set $\{v : c(v) = i\}$ is connected and $|\{v : c(v) = j\}| \leq 1$''), then a \emph{generalized {\rLCproblem{r}}} consists of finding the minimum weight (according to the order $\wlesseq$) of a proper color assignment $c$ that satisfies the properties in $\globalPropSet(c)$.

%-------------------------------------------------------------------------------------------------------------------------------------------------
% Examples
\subsection{Examples}
\label{sec:problemDescriptionEx}
Many different optimization and decision problems can be modeled as {\rLCproblems{r}} (for a decision problem we can say, for example, that the answer is ``no'' if and only if the minimum weight of a proper coloring is $\Error$).

For the examples shown throughout this paper, we will assume that, otherwise stated, the definitions of $L_v$ are for all $v \in V(G)$, of $\textsc{w}_{v,i}$ for all $v \in V(G)$ and all $i \in L_v$, and of $check(v,c)$ for all $v \in V(G)$ and all color assignments $c$ valid in $N_G[v]$. Also, if the labels $\ell_{e}$ are not specified, we can assume they are all equal to $1$.

In Table \ref{tab:examples} we show some examples of coloring, domination, independence and packing problems as {\rLCproblems{1}}. Their definitions can be found in Appendix~\ref{sec:problemsDef}.

%------------
{\renewcommand{\arraystretch}{1.5}\footnotesize
%--------------------------------------------------------------------------------------------------------------------------------------------------
%--------------------------------------------------------------------------------------------------------------------------------------------------
\begin{table}
    \centering
    \begin{tabular}{|c|c|c|c|c|c|c|c|c|c|} \hline
        Problem & $\WeightSet, \wlesseq, \weightsSum$ & $L_v$ & $\textsc{w}_{v,i}$ & $check(v,c)$ \\ \hline \hline
        %--------------------------------------------------------------------------------------------------------------------------------
        $k$-coloring
        & $\mathbb{N} \cup \{+\infty\}, \leq, \max$
        & $\intInterval{1}{k}$
        & $i$
        & $\forallFormula{u \in N_G(v)}{c(u) \neq c(v)}$ \\ \hline
        %---------------------
        $k$-chromatic sum
        & $\mathbb{N} \cup \{+\infty\}, \leq, +$
        & $\intInterval{1}{k}$
        & $i$
        & $\forallFormula{u \in N_G(v)}{c(u) \neq c(v)}$ \\ \hline
        %---------------------
        List-coloring
        & $\mathbb{N} \cup \{+\infty\}, \leq, +$
        & input
        & $0$
        & $\forallFormula{u \in N_G(v)}{c(u) \neq c(v)}$ \\ \hline
        %---------------------
        $H$-coloring
        & $\mathbb{N} \cup \{+\infty\}, \leq, +$
        & $V(H)$
        & $0$
        & $\forallFormula{u \in N_G(v)}{c(u) \in N_H(c(v))}$ \\ \hline
        %--------------------------------------------------------------------------------------------------------------------------------
        $k$-tuple domination
        & $\mathbb{N} \cup \{+\infty\}, \leq, +$
        & $\{0, 1\}$
        & $i$
        & $\sum_{u \in N_G[v]} c(u) \geq k$ \\ \hline
        %---------------------
        Total $k$-tuple domination
        & $\mathbb{N} \cup \{+\infty\}, \leq, +$
        & $\{0, 1\}$
        & $i$
        & $\sum_{u \in N_G(v)} c(u) \geq k$ \\ \hline
        %---------------------
        $k$-domination
        & $\mathbb{N} \cup \{+\infty\}, \leq, +$
        & $\{0, 1\}$
        & $i$
        & $c(v) = 0 \Rightarrow \sum_{u \in N_G(v)} c(u) \geq k$ \\ \hline
        %---------------------
        $\{k\}$-domination
        & $\mathbb{N} \cup \{+\infty\}, \leq, +$
        & $\intInterval{0}{k}$
        & $i$
        & $\sum_{u \in N_G[v]} c(u) \geq k$ \\ \hline
        %---------------------
        $k$-rainbow domination
        & $\mathbb{N} \cup \{+\infty\}, \leq, +$
        & $2^{\intInterval{1}{k}}$
        & $|i|$
        & $\left|\bigcup_{u \in N_G[v]} c(u)\right| = k$ \\ \hline
        %---------------------
        Roman domination
        & $\mathbb{N} \cup \{+\infty\}, \leq, +$
        & $\{0, 1, 2\}$
        & $i$
        & $c(v) = 0 \Rightarrow \existsFormula{u \in N_G(v)}{c(u) = 2}$ \\ \hline
        %--------------------------------------------------------------------------------------------------------------------------------
        Independent set
        & $\mathbb{N} \cup \{-\infty\}, \geq, +$
        & $\{0, 1\}$
        & $i$
        & $c(v) = 1 \Rightarrow \sum_{u \in N_G(v)} c(u) = 0$ \\ \hline
        %---------------------
        $\{k\}$-packing function
        & $\mathbb{N} \cup \{-\infty\}, \geq, +$
        & $\intInterval{0}{k}$
        & $i$
        & $\sum_{u \in N_G[v]} c(u) \leq k$ \\ \hline
        %---------------------
        $\{k\}$-limited packing
        & $\mathbb{N} \cup \{-\infty\}, \geq, +$
        & $\{0,1\}$
        & $i$
        & $\sum_{u \in N_G[v]} c(u) \leq k$ \\ \hline
        %--------------------------------------------------------------------------------------------------------------------------------
    \end{tabular}
    \caption{Examples of {\rLCproblems{1}}.}
    \label{tab:examples}
\end{table}
%--------------------------------------------------------------------------------------------------------------------------------------------------
%--------------------------------------------------------------------------------------------------------------------------------------------------
}

%------------

Observe that for list-coloring and $H$-coloring we are only interested in determining whether such coloring exists or not, so we do not use the weights for optimizing the solution, instead we use them precisely for determining if a solution is correct. For $k$-coloring we can make use of weights to determine the smallest $j \leq k$ for which there exists a $j$-coloring in $G$ (in particular, we could set $k$ as a known upper bound for the chromatic number to obtain it).

For the total version of most of these domination problems we only need to replace $N[v]$ by $N(v)$.

For weighted versions of these problems, weights are part of the input.

For problems where the input graph is directed, we can consider edge labels that also carry the direction of the edge, and the $check$ function can use it to distinguish in-neighbors and out-neighbors.

%-------------------------------------------------------------------------------------------------------------------------------------------------
% LCVP
\subsection{{\LCVP} problems}
\label{sec:problemDescriptionLCVP}
The problem of deciding if a graph $G$ has a $D_q$ partition can be modeled as a {\rLCproblem{1}} in the following way:
\begin{itemize}
\item $L_{v} = \intInterval{1}{q}$;
\item $(\WeightSet, \wlesseq, \weightsSum) = (\mathbb{N} \cup \{+\infty\}, \leq, +)$;
\item $\textsc{w}_{v, i} = 0$; and
\item $check(v, c) = \forallFormula{j \in \intInterval{1}{q}}{|\setst{u}{u \in N(v) \land c(u) = j}| \in D_q[c(v),j]}$.
\end{itemize}
Therefore, our framework is a generalization of {\LCVP} problems. Furthermore, it allows us to model more problems, like $\{k\}$--domination.

%-------------------------------------------------------------------------------------------------------------------------------------------------
% LCC and ECC
\subsection{{\LCC} and {\ECC} problems}
\label{sec:problemDescriptionLCCECC}
Consider an {\ECC} problem $\Pi$. By definition, there exists a basic {\ECC} problem $\Pi'$ such that $\Pi$ can be reduced to $\Pi'$. We will show that $\Pi'$ can be reduced to a generalized {\rLCproblem{1}} in the jagged graph of the input graph.

Construct $J(G)$ from the input graph $G$.
Let $(M^{*}, \oplus^{*}, \preceq^{*})$ be obtained by adding a maximum element to $(M^{m+1}, \oplus^{m+1}, \preceq)$.
Then
\begin{itemize}
\item $L_{v} = X$ for all $v \in V(G)$, and\\
$L_{uv} = L_u \times L_v \times Y$ for all $uv \in E(G)$;
\item $(\WeightSet, \wlesseq, \weightsSum) = (M^{*}, \preceq^{*}, \oplus^{*})$;
\item $\textsc{w}_{v, x} = \weightsNeutral$ for all $v \in V(G)$, and\\
$\textsc{w}_{uv, (x_u, x_v, y)} = val_{m+1}(D, uv, x_u, x_v, y)$ for all $uv \in E(G)$;
\item $check(v, c) = \True$ for all $v \in V(G)$ and all color assignments $c$ valid in $N_{J(G)}[v]$, and\\
$check(uv, c) = (c(uv)_1 = c(u) \land c(uv)_2 = c(v))$ for all $uv \in E(G)$ and all color assignments $c$ valid in $N_{J(G)}[e]$; and
\item Global properties $\globalPropSet(c)$: $\bigoplus^i_{uv \in E(G)} val_i\left(D, uv, c(uv)_1, c(uv)_2, c(uv)_3\right) \in R_i$, for all $i \in \intInterval{1}{m}$.
\end{itemize}

With a similar approach, we can reduce a {\LCC} problem in bounded degree graphs to a generalized {\rLCproblem{r}} in the jagged graph of the input graph (for an appropriate $r$).
Construct $J(G)$ from the input graph $G$.
Let $(M^{*}, \oplus^{*}, \preceq^{*})$ be obtained by adding a maximum element to $(M^{m+1}, \oplus^{m+1}, \preceq)$.
For all $v \in V(G)$, let $\mathbb{F}_v$ and $\mathbb{G}_v$ be, respectively, the sets of all functions $f \colon N_G^r[v] \to X$ and $g \colon M_G^r(v) \to Y$ (notice that $\mathbb{F}_v$ and $\mathbb{G}_v$ are finite sets that can be computed in polynomial time).
Then
\begin{itemize}
\item $L_{v} = \mathbb{F}_v \times \mathbb{G}_v$ for all $v \in V(G)$, and\\
$L_{uv} = Y$ for all $uv \in E(G)$;
\item $(\WeightSet, \wlesseq, \weightsSum) = (M^{*}, \preceq^{*}, \oplus^{*})$;
\item $\textsc{w}_{v, (f,g)} = val_{m+1}(D, v, f, g)$ for all $v \in V(G)$, and\\
$\textsc{w}_{uv, y} = \weightsNeutral$ for all $uv \in E(G)$;
\item for all $v \in V(J(G))$ and all color assignments $c$ valid in $N_{J(G)}^r[v]$, $check(v, c)$ checks that $f_u(x) = f_w(x)$ and $g_u(y) = g_w(y)$ (where $(f_u,g_u) = c(u)$ and $(f_w, g_w) = c(w)$) for all $u,w \in N_{J(G)}^r[v] \cap V(G)$, $x \in N_{J(G)}^r[v] \cap N_G^r[u] \cap N_G^r[w]$ and $y \in N_{J(G)}^r[v] \cap M_G^r(u) \cap M_G^r(w)$; and
\item Global properties $\globalPropSet(c)$: $\bigoplus^i_{v \in V(G)} val_i\left(D, v, c(v)_1, c(v)_2\right) \in R_i$, for all $i \in \intInterval{1}{m}$.
\end{itemize}

\section{{\rLCproblems{1}} in complete graphs}
\label{sec:completeGraphs}

It is easy to see that we can polynomially reduce NP-complete problems in graphs to particular {\rLCproblems{1}} in complete graphs, even when restricting the sets of colors and edge labels to $\{0,1\}$. Indeed, we can transform the classical domination problem in a graph $G$ to the following {\rLCproblem{1}} in complete graphs. We construct a complete graph $G'$ such that $V(G') = V(G)$ and set
\begin{itemize}
\item $\ell_{uv} = 1$ if $uv \in E(G)$ and $\ell_{uv} = 0$ otherwise;
\item $L_{v} = \{0, 1\}$;
\item $(\WeightSet, \wlesseq, \weightsSum) = (\mathbb{R} \cup \{+\infty\}, \leq, +)$;
\item $\textsc{w}_{v, i} = i$; and
\item $check(v, c) = \left(c(v) + \sum_{u \in N_{G'}(v)}(c(u) \cdot \ell_{vu}) \geq 1\right)$.
\end{itemize}

It is clear that the minimum weight of a proper coloring in this instance equals $\gamma(G)$, and this transformation can be performed in polynomial time.

If we restrict $\LabelSet$ to $\{1\}$, we can still find a polynomial-time reduction from the domination problem in a graph $G$ to a {\rLCproblem{1}} in a complete graph $G'$ such that $V(G') = V(G)$, by setting:
\begin{itemize}
\item $L_{v} = \{\emptyset, N_G[v]\}$;
\item $(\WeightSet, \wlesseq, \weightsSum) = (\mathbb{R} \cup \{+\infty\}, \leq, +)$;
\item $\textsc{w}_{v, \emptyset} = 0$ and $\textsc{w}_{v, N_G[v]} = 1$; and
\item $check(v, c) = \left(v \in \bigcup_{u \in N_{G'}[v]}c(u)\right)$.
\end{itemize}

Of course, {\rLCproblems{1}} are polynomial-time solvable under appropriate conditions.
\begin{theorem}
Consider a {\rLCproblem{1}} and a family of instances where
\begin{itemize}
\item $G$ is a complete graph;
\item $\ell_{e} = 1$ for all $e \in E(G)$;
\item the number of all possible colors (that is, the size of the set $\bigcup_{v \in V(G)} L_v$) is bounded by a constant; and
\item $check(v, c)$ can be computed in polynomial time and only depends on $v$, $c(v)$ and the number of neighbors of each color that $v$ has.
\end{itemize}
Then this problem can be solved in polynomial time in $|V(G)|$ for these instances.
\begin{proof}
Assume $\textsc{Colors} = \{c_1, \ldots, c_{\mathcal{C}}\}$.
Notice that since $G$ is a complete graph then $N_{G}[v] = V(G)$ for all $v \in V(G)$.
Therefore, by the restrictions imposed to $check$, we can assume there exists a function $check'$ such that $check'(v, c(v), (k_1, \ldots, k_{\mathcal{C}})) = check(v,c)$, where $k_i = |\setst{u \in V(G)}{c(u) = c_i}|$ for all $i \in \intInterval{1}{\mathcal{C}}$.

For every distribution of colors $(k_1, \ldots, k_{\mathcal{C}})$, with $k_i \in \mathbb{N}_0$ for all $i \in \intInterval{1}{\mathcal{C}}$ and such that $\sum_{i=1}^{\mathcal{C}} k_i = |V(G)|$, we need to verify if it can actually be achieved (that is, there exists a proper color assignment $c$ such that $k_i = |\setst{u \in V(G)}{c(u) = c_i}|$ for all $i \in \intInterval{1}{\mathcal{C}}$), and if so, find one such proper assignment of colors to vertices of minimum weight. To this end, we construct a directed capacitated network $F$ with vertices $i$ for all $i \in \textsc{Colors}$, $(v,i)$ for all $v \in V(G)$ and all $i \in L_v$, $v$ for all $v \in V(G)$, and $s$ and $t$.
There is a directed edge of capacity $k_{\sigma(i)}$ and cost $0$ from $s$ to $i$ for all $i \in \textsc{Colors}$.
There is a directed edge of capacity $1$ and cost $0$ from $i$ to $(v,i)$ if $k_i \geq 1$ and $check'(v, i, (k_1, \ldots, k_{\mathcal{C}}))$ is true.
There is a directed edge of capacity $1$ and cost $\textsc{w}_{v,i}$ from $(v,i)$ to $v$ for all $v \in V(G), i \in L_v$.
There is a directed edge of capacity $1$ and cost $0$ from $v$ to $t$ for all $v \in V(G)$.
There are no more edges than these ones. Note that the distribution $(k_1, \ldots, k_{\mathcal{C}})$ is achievable if and only if the maximum flow in $F$ is $|V(G)|$, and in this case the proper assignment of colors to vertices of minimum weight corresponds to the maximum flow in $F$ of minimum cost.

Finally, the answer to the problem is obtained by finding the minimum proper color assignment among the ones found for all achievable distributions of colors.

Since $\mathcal{C}$ is bounded by a constant, the number of distributions of colors is polynomial in $|V(G)|$ (because it is $\binom{|V(G)|+\mathcal{C}-1}{\mathcal{C}-1} < (|V(G)|+1)^{\mathcal{C}}$), $check'(v, i, (k_1, \ldots, k_{\mathcal{C}}))$ can be computed in polynomial time, constructing $F$ takes polynomial time, and the problem of finding the maximum flow of minimum cost is polynomial-time solvable, then the statement holds.
\end{proof}
\end{theorem}

\section{{\rLCproblems{1}} in bounded treewidth graphs}
\label{sec:boundedTreewidth}

In this section we give a polynomial-time algorithm to solve {\rLCproblems{1}} in bounded treewidth graphs, under mild conditions.
In Section \ref{sec:boundedDegree} we show that these results can be extended to {\rLCproblems{r}} (for any fixed $r \geq 1$) in bounded treewidth and bounded degree graphs.
In Section \ref{sec:globalProp} we will explain how to modify the algorithm in order to add some global properties.

We will solve the problem in a dynamic programming fashion, traversing an easy tree decomposition (see Definition \ref{def:easytreedec}) of the input graph and describing how to proceed with each type of node of the tree decomposition, as it is usual with this kind of problems, but with an abstraction of the ``extra'' parameters involved in ad-hoc solutions of locally checkable problems. In order to describe this abstraction, and hence the algorithm, we first need to introduce the concept of \emph{partial neighborhoods}.

%---------------------------------------------------------------------------------------------------
\subsection{Partial neighborhoods}\label{sec:partialNeighborhoods}
We define a system that, roughly speaking, gives us tools to accumulate information from the neighbors of a vertex.

\begin{definition}
A \emph{partial neighborhood system} for an instance of a {\rLCproblem{1}} consists of:
\begin{itemize}
\item A set $\NeighborhoodSet{v}{i}$, for every $v \in V(G)$ and $i \in L_{v}$, together with a closed binary operation $\neighborhoodSum{v}{i}$ on $\NeighborhoodSet{v}{i}$ that is commutative and associative and has a neutral element $\emptyNeighborhood{v}{i}$.

\item A function $\toNeighborhood{v}{i}$, for every $v \in V(G)$ and $i \in L_{v}$, that given $u \in N_G(v)$ and $j \in L_u$ returns an element of $\NeighborhoodSet{v}{i}$ (possibly making use of the label of the edge $vu$).

\item A function $check_{v,i} \colon \NeighborhoodSet{v}{i} \to \BoolSet$, for every $v \in V(G)$ and $i \in L_{v}$. This function must satisfy
$check_{v,c(v)}\left(\bigneighborhoodSumSym^{v,c(v)}_{u \in N_G(v)} \toNeighborhood{v}{c(v)}(u, c(u))\right) = check(v,c)$
for every vertex $v \in V(G)$ and every color assignment $c$ valid in $N_G[v]$.
\end{itemize}
\end{definition}

In words, with $\toNeighborhood{v}{i}(u,j)$ we create new information, that says how $u$ having color $j$ affects $v$ when having color $i$. The operation $\neighborhoodSum{v}{i}$ combines two pieces of information. For a color assignment $c$ valid in $N_G[v]$, $check_{v,c(v)}\left(\bigneighborhoodSumSym^{v,c(v)}_{u \in N_G(v)} \toNeighborhood{v}{c(v)}(u, c(u))\right)$ simply verifies a condition over all the information collected from the neighbors of $v$. Finally, we require the equality $check_{v,c(v)}\left(\bigneighborhoodSumSym^{v,c(v)}_{u \in N_G(v)} \toNeighborhood{v}{c(v)}(u, c(u))\right) = check(v,c)$ to make these tools analogous to the use of $check(v,c)$. We refer to the elements of $\NeighborhoodSet{v}{i}$ as \emph{partial neighborhoods} of vertex $v$ with color $i$.

\begin{remark}\label{vectorPartialNeighborhood}
For every instance of a {\rLCproblem{1}} there exists a partial neighborhood system.
We will show how to construct one.
The idea behind the following partial neighborhood system is to store all the colors assigned to the neighbors of $v$, where $\perp$ represents that a neighbor has not yet been assigned a color, and $\times$ can be thought as an error sign.
Let $v \in V(G)$, $i \in L_v$ and assume $N_G(v) = \{u_1, \ldots, u_{d_G(v)}\}$.
Let $\NeighborhoodSet{v}{i}$ be the set of all $d_G(v)$-tuples $x$ such that $x_{h} \in L_{u_h} \cup \{\perp, \times\}$ for all $h \in \intInterval{1}{d_G(v)}$.
Let $\neighborhoodSum{v}{i}$ be such that
$$(n \neighborhoodSum{v}{i} n')_{h} = \begin{cases}
n_{h}	&	\text{if } n_{h} = n'_{h} \text{ or } n'_{h} = \perp\\
n'_{h}	&	\text{if } n_{h} = \perp \text{ and } n'_{h} \neq \perp\\
\times	&	\text{otherwise}
\end{cases}$$
for all $h \in \intInterval{1}{d_G(v)}$.
Let $\toNeighborhood{v}{i}(u_h,j)$ be the $d_G(v)$-tuple that has $j$ in its position $h$ and $\perp$ in all its other positions.
Let $x \in \NeighborhoodSet{v}{i}$. If $x = (j_1, \ldots, j_{d_G(v)})$ with $j_h \in L_{u_h}$ for all $h \in \intInterval{1}{d_G(v)}$, let $c$ be the color assignment in $N_G[v]$ such that $c(v) = i$ and $c(u_h) = j_h$ for all $h \in \intInterval{1}{d_G(v)}$, and then define $check_{v,i}(x) = check(v, c)$. Otherwise, let $check_{v,i}(x) = \False$.
\end{remark}

{\renewcommand{\arraystretch}{1.5}\footnotesize
%--------------------------------------------------------------------------------------------------------------------------------------------------
%--------------------------------------------------------------------------------------------------------------------------------------------------
\begin{table}
    \centering
    \begin{tabular}{|c|c|c|c|c|c|c|c|c|c|} \hline
        Problem & $\NeighborhoodSet{v}{i}$ & $n \neighborhoodSum{v}{i} n'$ & $\toNeighborhood{v}{i}(u, j)$ & $check_{v,i}(n)$ \\ \hline \hline
        %--------------------------------------------------------------------------------------------------------------------------------
        $k$-coloring
        & $\BoolSet$
        & $n \land n'$
        & $j \neq i$
        & $n$ \\ \hline
        %---------------------
        $k$-chromatic sum
        & $\BoolSet$
        & $n \land n'$
        & $j \neq i$
        & $n$ \\ \hline
        %---------------------
        List-coloring
        & $\BoolSet$
        & $n \land n'$
        & $j \neq i$
        & $n$ \\ \hline
        %---------------------
        $H$-coloring
        & $\BoolSet$
        & $n \land n'$
        & $j \in N_H(i)$
        & $n$ \\ \hline
        %--------------------------------------------------------------------------------------------------------------------------------
        $k$-tuple domination
        & $\intInterval{0}{k}$
        & $\min(n + n', k)$
        & $j$
        & $n + i \geq k$ \\ \hline
        %---------------------
        Total $k$-tuple domination
        & $\intInterval{0}{k}$
        & $\min(n + n', k)$
        & $j$
        & $n \geq k$ \\ \hline
        %---------------------
        $k$-domination
        & $\intInterval{0}{k}$
        & $\min(n + n', k)$
        & $j$
        & $i = 0 \Rightarrow n \geq k$ \\ \hline
        %---------------------
        $\{k\}$-domination
        & $\intInterval{0}{k}$
        & $\min(n + n', k)$
        & $j$
        & $n + i \geq k$ \\ \hline
        %---------------------
        $k$-rainbow domination
        & $2^{\intInterval{1}{k}}$
        & $n \cup n'$
        & $j$
        & $\left|n \cup i \right| = k$ \\ \hline
        %---------------------
        Roman domination
        & $\BoolSet$
        & $n \lor n'$
        & $j = 2$
        & $i = 0 \Rightarrow n$ \\ \hline
        %--------------------------------------------------------------------------------------------------------------------------------
        Independent set
        & $\{0,1\}$
        & $\min(n + n', 1)$
        & $j$
        & $i = 1 \Rightarrow n = 0$ \\ \hline
        %---------------------
        $\{k\}$-packing function
        & $\intInterval{0}{k+1}$
        & $\min(n + n', k+1)$
        & $j$
        & $n \leq k$ \\ \hline
        %---------------------
        $\{k\}$-limited packing
        & $\intInterval{0}{k+1}$
        & $\min(n + n', k+1)$
        & $j$
        & $n \leq k$ \\ \hline
        %--------------------------------------------------------------------------------------------------------------------------------
    \end{tabular}
    \caption{Examples of partial neighborhood systems for {\rLCproblems{1}}.}
    \label{tab:partialNeighborhoods}
\end{table}
%--------------------------------------------------------------------------------------------------------------------------------------------------
%--------------------------------------------------------------------------------------------------------------------------------------------------
}

Finding partial neighborhood systems that have smaller sets $\NeighborhoodSet{v}{i}$ is of extreme importance because it reduces the time complexity of the algorithm given in Section \ref{subsec:alg}.
We say a partial neighborhood system is \emph{\pnPolynomial} (resp. \emph{\pnConstant}) if it is such that $\max\setst{|\NeighborhoodSet{v}{i}|}{v \in V(G), i \in L_v}$ is polynomial (resp. constant) in the size of the input of the problem, and all the functions $check_{v,i}$, $eq_{n}$, $\neighborhoodSum{v}{i}$ and $\toNeighborhood{v}{i}$ can be computed in time polynomial (resp. constant) in the input size.
In Table~\ref{tab:partialNeighborhoods} we show {\pnConstant} partial neighborhood systems of the locally checkable problems listed in Table~\ref{tab:examples}.

Another key concept is the following.
Let $X \subseteq V(G)$ and $c$ be a color assignment valid in $X$. Given a partial neighborhood system, a \emph{valid partial neighborhood mapping for $c$} is a function $\eta$ of domain $X$ such that $\eta(v) \in \NeighborhoodSet{v}{c(v)}$ for all $v \in X$.

Finally, given $v \in V(G)$ and $i \in L_v$, any function $f \colon \NeighborhoodSet{v}{i} \to \BoolSet$ is called a \emph{checking function for $(v, i)$}.

%---------------------------------------------------------------------------------------------------
\subsection{Notation and definitions}
The following definitions and notation will be useful throughout the rest of the article. To make the notation less cumbersome, we write $\checkFuns_v$ instead of $\checkFuns(v)$.

\begin{itemize}
%---------------------------------------------------------------------------------------------------
\item \textbf{Function extended with one element in its domain.} Let $f \colon X \rightarrow Y$ and $x \notin X$. Then the function $f^{x \rightarrow y} \colon X \cup \{x\} \rightarrow Y \cup \{y\}$ is such that $f^{x \rightarrow y}(x) = y$ and $f^{x \rightarrow y}(z) = f(z)$ for all $z \in X$.

%---------------------------------------------------------------------------------------------------
\item \textbf{Function with one element less in its domain.} Let $f \colon X \rightarrow Y$ and $x \in X$. Then the function $f^{-x} \colon X - \{x\} \rightarrow Y$ is such that $f^{-x}(z) = f(z)$ for all $z \in X - \{x\}$.

%---------------------------------------------------------------------------------------------------
\item \textbf{Graph where some edges are removed.} Let $H$ be a graph. Then $\removeEdgesSet{H}{S}$ is the graph such that $V(\removeEdgesSet{H}{S}) = V(H)$ and $E(\removeEdgesSet{H}{S}) = E(H) - \setst{uv}{u,v \in S}$.

%---------------------------------------------------------------------------------------------------
\item \textbf{Neutral weight mapping.} Let $X \subseteq V(G)$. Then the function $\weightsFun^e_X \colon X \to \WeightSet$ is such that $\weightsFun^e_X(v) = \weightsNeutral$ for all $v \in X$.

%---------------------------------------------------------------------------------------------------
\item \textbf{Equality checking function.} For every $v \in V(G)$, every $i \in L_v$ and every $n \in \NeighborhoodSet{v}{i}$, let $eq_{n} \colon \NeighborhoodSet{v}{i} \rightarrow \BoolSet$ be the function such that $eq_{n}(n') = (n = n')$ for all $n' \in \NeighborhoodSet{v}{i}$.

%---------------------------------------------------------------------------------------------------
\item \textbf{Function that returns equality checking functions.} Let $X \subseteq V(G)$, $c$ be a color assignment valid in $X$ and $\eta$ be a valid partial neighborhood mapping for $c$. Then let $\checkFuns^{\eta}$ be the function of domain $X$ such that $\checkFuns^{\eta}_v$ is $eq_{\eta(v)}$ for all $v \in X$.

%---------------------------------------------------------------------------------------------------
\item \textbf{Reduction of a partial neighborhood mapping.} Let $X \subseteq V(G)$, $c$ be a color assignment valid in $X$, $\eta$ be a valid partial neighborhood mapping for $c$ and $v \in X$. Then the function $\eta^{\sim v}$ of domain $X - \{v\}$ is such that $\eta^{\sim v}(u) = \eta(u) \neighborhoodSum{u}{c(u)} \toNeighborhoodSym(v, c(v))$ if $u \in X \cap N_G(v)$ and $\eta^{\sim v}(u) = \eta(u)$ otherwise.

%---------------------------------------------------------------------------------------------------
\item \textbf{Neutral partial neighborhood mapping.} Let $X \subseteq V(G)$ and $c$ be a color assignment valid in $X$. Then the function $\eta^e_c$ of domain $X$ is such that $\eta^e_c(v) = \emptyNeighborhood{v}{c(v)}$ for all $v \in X$. Observe that $\eta^e_c$ is a valid partial neighborhood mapping for $c$.

%---------------------------------------------------------------------------------------------------
\item \textbf{Partial neighborhood in a subgraph.} Let $H$ be a subgraph of $G$ and $c$ be a color assignment valid in $V(H)$. Then we define $\pnsubg$ such that $\pnsubg(v,c,H) = \bigneighborhoodSumSym^{v,c(v)}_{u \in N_{H}(v)} \toNeighborhood{v}{c(v)}(u, c(u))$ for all $v \in V(H)$. Roughly speaking, $\pnsubg(v,c,H)$ is the information we can obtain from the neighbors of $v$ in $H$ and the color assignment $c$.
\end{itemize}

%---------------------------------------------------------------------------------------------------
\subsection{Algorithm}
\label{subsec:alg}
Consider an instance of a {\rLCproblem{1}} with a partial neighborhood system. Let $G$ be the input graph and let $(T, \{X_t\}_{t \in V(T)})$ be an easy tree decomposition of $G$.

%---------------------------------------------------------------------------------------------------
For every $X \subseteq V(G)$,
every $G'$ subgraph of $G$ such that $X \subseteq V(G')$ and $N_{G}[V(G') - X] \subseteq V(G')$,
every color assignment $c$ valid in $X$,
every $\eta$ that is a valid partial neighborhood mapping for $c$,
and every $\checkFuns$ such that $\checkFuns_v \in \{check_{v,c(v)}\} \cup \setst{eq_n}{n \in \NeighborhoodSet{v}{c(v)}}$ for all $v \in X$,
we say that a function $f$ is a \emph{$(X, c, \eta, \checkFuns)$--coloring in $G'$} if
\begin{itemize}
\item $f$ is a color assignment valid in $V(G')$,
\item $f(v) = c(v)$ for all $v \in X$,
\item $\checkFuns_v\left(\eta(v) \neighborhoodSum{v}{f(v)} \pnsubg(v,f,G')\right) = \True$ for all $v \in X$, and
\item $check(u, f|_{N_{G}[u]}) = \True$ for all $u \in V(G') - X$.
\end{itemize}

%---------------------------------------------------------------------------------------------------
For a $(X, c, \eta, \checkFuns)$--coloring $f$ in $G'$ and a function $\weightsFun \colon X \to \WeightSet$, we define the \emph{weight under $\weightsFun$ of $f$} as $\textsc{w}_{\weightsFun}(f) = \left(\bigweightsSum_{v \in X} \weightsFun(v)\right) \weightsSum \left(\bigweightsSum_{v \in V(G') - X} \textsc{w}_{v, f(v)}\right)$.

%---------------------------------------------------------------------------------------------------
For every node $t$, let $\genDomDomain{t}$ be the set of all tuples $(S, c, \weightsFun, \eta, \checkFuns)$ such that
\begin{itemize}
\item $S \subseteq X_t$,
\item $c$ is a color assignment valid in $X_t$,
\item $\weightsFun \colon X_t \to \WeightSet$ is such that $\weightsFun(v) \in \{\weightsNeutral, \textsc{w}_{v,c(v)}\}$ for all $v \in X_t$,
\item $\eta$ is a valid partial neighborhood mapping for $c$, and
\item $\checkFuns$ is a function that, given a vertex $v \in X_t$, returns a \emph{checking function for $(v, c(v))$} such that $\checkFuns_v \in \{check_{v,c(c)}\} \cup \setst{eq_{n}}{n \in \NeighborhoodSet{v}{c(v)}}$;
\end{itemize}
and let $\genDomSym_t \colon \genDomDomain{t} \to \WeightSet$ be defined by
$$\genDomSym_t(S, c, \weightsFun, \eta, \checkFuns) = \minWeights\setst{\textsc{w}_{\weightsFun}(f)}{\text{$f$ is a $(X_t, c, \eta, \checkFuns)$--coloring in $\removeEdgesSet{G_t}{S}$}}$$
for every $(S, c, \weightsFun, \eta, \checkFuns) \in \genDomDomain{t}$.

%---------------------------------------------------------------------------------------------------
\begin{remark}
Notice that if there are no $(X_t, c, \eta, \checkFuns)$--colorings in $\removeEdgesSet{G_t}{S}$ then we have $\genDomSym_t(S, c, \weightsFun, \eta, \checkFuns) = \Error$.
\end{remark}

%---------------------------------------------------------------------------------------------------
The following result is immediate from the previous definitions.
\begin{corollary}
If $r$ is the root of $T$ then the minimum weight of a proper coloring in $G$ is
\begin{align*}
\minWeights\{
\genDomSym_r(\emptyset, c, \weightsFun, \eta^e_c, \checkFuns)
:\;
&(\emptyset, c, \weightsFun, \eta^e_c, \checkFuns) \in \genDomDomain{r}, \text{and}\\
&\text{ $\weightsFun(v) = \textsc{w}_{v, c(v)}$ and $\checkFuns_v = check_{v,c(v)}$ for all $v \in X_r$}
\}.
\end{align*}
\end{corollary}

%--------------------------------------------------------------------------------------------------------------------------------------------------
%--------------------------------------------------------------------------------------------------------------------------------------------------
We will show how to compute $\genDomSym$ in a recursive way.

%-----------------------------
% Leaf
\begin{lemma}[Leaf node]\label{lem:leaf}
Let $t$ be a leaf node and $X_t = \{v\}$. Then
$$\genDomSym_t(S, c, \weightsFun, \eta, \checkFuns) = \left\{\begin{array}{ll}
\weightsFun(v) & \text{if } \checkFuns_v(\eta(v)) \\
\Error  & \text{otherwise}
\end{array}\right.$$
\begin{proof}
Notice that in this case $V(\removeEdgesSet{G_t}{S}) = \{v\} = X_t$.

If $\checkFuns_v(\eta(v)) = \False$ then, by definition, there are no $(X_t, c, \eta, \checkFuns)$--colorings in $\removeEdgesSet{G_t}{S}$. Therefore, $\genDomSym_t(S, c, \weightsFun, \eta, \checkFuns) = \Error$, leading to the desired equality.

If $\checkFuns_v(\eta(v)) = \True$ then, by definition, $c$ is the only possible $(X_t, c, \eta, \checkFuns)$--coloring in $\removeEdgesSet{G_t}{S}$. Therefore,
$\genDomSym_t(S, c, \weightsFun, \eta, \checkFuns) = \minWeights\setst{\textsc{w}_{\weightsFun}(f)}{\text{$f$ is a $(X_t, c, \eta, \checkFuns)$--coloring in $\removeEdgesSet{G_t}{S}$}} = \textsc{w}_{\weightsFun}(c) = \weightsFun(v)$.
\end{proof}
\end{lemma}

%-----------------------------
% Forget
\begin{lemma}[Forget node]\label{lem:forget}
Let $t$ be a forget node, $s$ be the child of $t$ and $X_s - X_t = \{v\}$. Then
$$\genDomSym_t(S, c, \weightsFun, \eta, \checkFuns) = \minWeights_{i \in L_v}\{\genDomSym_s(S, c^{v \rightarrow i}, \weightsFun^{v \rightarrow \textsc{w}_{v,i}}, \eta^{v \rightarrow \emptyNeighborhood{v}{i}}, \checkFuns^{v \rightarrow check_{v,i}})\}.$$
\begin{proof}
Notice that $v \notin S$ and also $\removeEdgesSet{G_t}{S} = \removeEdgesSet{G_s}{S}$.

By definition of $\genDomSym_s$ we have
\begin{align*}
	&\minWeights_{i \in L_v}\{\genDomSym_s(S, c^{v \rightarrow i}, \weightsFun^{v \rightarrow \textsc{w}_{v,i}}, \eta^{v \rightarrow \emptyNeighborhood{v}{i}}, \checkFuns^{v \rightarrow check_{v,i}})\}\\
=\; &\minWeights_{i \in L_v}\{\minWeights\setst{\textsc{w}_{\weightsFun^{v \rightarrow \textsc{w}_{v,i}}}(f)}{\text{$f$ is a $(X_s, c^{v \rightarrow i}, \eta^{v \rightarrow \emptyNeighborhood{v}{i}}, \checkFuns^{v \rightarrow check_{v,i}})$--coloring in $\removeEdgesSet{G_s}{S}$}}\}\\
=\; &\minWeights\setst{\textsc{w}_{\weightsFun^{v \rightarrow \textsc{w}_{v,i}}}(f)}{i \in L_v \land \text{$f$ is a $(X_s, c^{v \rightarrow i}, \eta^{v \rightarrow \emptyNeighborhood{v}{i}}, \checkFuns^{v \rightarrow check_{v,i}})$--coloring in $\removeEdgesSet{G_s}{S}$}}.
\end{align*}

Let $i \in L_v$. We claim that every $f$ that is a $(X_s, c^{v \rightarrow i}, \eta^{v \rightarrow \emptyNeighborhood{v}{i}}, \checkFuns^{v \rightarrow check_{v,i}})$--coloring in $\removeEdgesSet{G_s}{S}$ is also a $(X_t, c, \eta, \checkFuns)$--coloring in $\removeEdgesSet{G_t}{S}$. Conversely, every $f$ that is a $(X_t, c, \eta, \checkFuns)$--coloring in $\removeEdgesSet{G_t}{S}$ is also a $(X_s, c^{v \rightarrow f(v)}, \eta^{v \rightarrow \emptyNeighborhood{v}{f(v)}}, \checkFuns^{v \rightarrow check_{v,f(v)}})$--coloring in $\removeEdgesSet{G_s}{S}$.
We prove the first claim (the second one is similar) by showing each of the items of the definition of $(X_t, c, \eta, \checkFuns)$--coloring in $\removeEdgesSet{G_t}{S}$ holds:
	\begin{itemize}
	\item $f(w) = c(w)$ for all $w \in X_t$ (because it is true for all $w \in X_s$),
	\item $f$ is a color assignment valid in $V(\removeEdgesSet{G_s}{S}) = V(\removeEdgesSet{G_t}{S})$,
	\item $check(u, f) = \True$ for all $u \in V(\removeEdgesSet{G_s}{S}) - X_s = V(\removeEdgesSet{G_t}{S}) - X_t - \{v\}$ and\\
		$\begin{aligned}[t]
		check(v, f) &= check_{v,i}\left(\pnsubg(v,f,\removeEdgesSet{G_s}{S})\right)\\
				&= check_{v,i}\left(\emptyNeighborhood{v}{i} \neighborhoodSum{v}{i} \pnsubg(v,f,\removeEdgesSet{G_s}{S})\right)\\
				&= \checkFuns^{v \rightarrow check_{v,i}}_v\left(\eta(v) \neighborhoodSum{v}{i} \pnsubg(v,f,\removeEdgesSet{G_s}{S})\right)\\
				&= \True
	\end{aligned}$\\
	(because $f(v) = i$), and
	\item $\checkFuns_{w}\left(\eta(w) \neighborhoodSum{w}{f(w)} \pnsubg(w,f,\removeEdgesSet{G_t}{S}) \right) = \True$ for all $w \in X_t$ (because it is true for all $w \in X_s$ and $E(\removeEdgesSet{G_t}{S}) = E(\removeEdgesSet{G_s}{S})$).
	\end{itemize}

Clearly, $\textsc{w}_{\weightsFun^{v \rightarrow \textsc{w}_{v,f(v)}}}(f) = \textsc{w}_{\weightsFun}(f)$ for every $f$ that is a $(X_t, c, \eta, \checkFuns)$--coloring in $\removeEdgesSet{G_t}{S}$.
Therefore
\begin{align*}
	&\minWeights\setst{\textsc{w}_{\weightsFun^{v \rightarrow \textsc{w}_{v,i}}}(f)}{i \in L_v \land \text{$f$ is a $(X_s, c^{v \rightarrow i}, \eta^{v \rightarrow \emptyNeighborhood{v}{i}}, \checkFuns^{v \rightarrow check_{v,i}})$--coloring in $\removeEdgesSet{G_s}{S}$}}\\
=\;	&\minWeights\setst{\textsc{w}_{\weightsFun}(f)}{\text{$f$ is a $(X_t, c, \eta, \checkFuns)$--coloring in $\removeEdgesSet{G_t}{S}$}}\\
=\;	&\genDomSym_t(S, c, \weightsFun, \eta, \checkFuns).
\end{align*}
%as we wanted.
\end{proof}
\end{lemma}

%-----------------------------
% Introduce
\begin{lemma}[Introduce node]\label{lem:introduce}
Let $t$ be an introduce node, $s$ be the child of $t$ and $X_t - X_s = \{v\}$. Let $n_v = \eta(v) \neighborhoodSum{v}{c(v)} \pnsubg(v,c,\removeEdgesSet{G_t}{S}[X_t])$. Then
$$\genDomSym_t(S, c, \weightsFun, \eta, \checkFuns) = \left\{\begin{array}{ll}
\weightsFun(v) \weightsSum \genDomSym_s(S - \{v\}, c^{-v}, \weightsFun^{-v}, \eta^{\sim v}, \checkFuns^{-v}) & \text{if } \checkFuns_v(n_v) \\
\Error & \text{otherwise}
\end{array}\right.$$
\begin{proof}
Observe that $v \notin V(\removeEdgesSet{G_s}{(S - \{v\})})$ and $N_{\removeEdgesSet{G_t}{S}}[v] \subseteq X_t$ (because $(T, \{X_i\}_{i \in V(T)})$ is a tree-decomposition), and that this implies that $\pnsubg(v,c,\removeEdgesSet{G_t}{S}) = \pnsubg(v,c,\removeEdgesSet{G_t}{S}[X_t])$.

If $\checkFuns_{v}(n_v) = \False$ then, by definition, there are no $(X_t, c, \eta, \checkFuns)$--colorings in $\removeEdgesSet{G_t}{S}$ and we also have $\genDomSym_t(S, c, \weightsFun, \eta, \checkFuns) = \Error$.

Now assume that $\checkFuns_{v}(n_v) = \True$.
It is straightforward to prove that if a function $f$ is a $(X_s, c^{-v}, \eta^{\sim v}, \checkFuns^{-v})$--coloring in $\removeEdgesSet{G_s}{(S - \{v\})}$ then the function $f^{v \rightarrow c(v)}$ is a $(X_t, c, \eta, \checkFuns)$--coloring in $\removeEdgesSet{G_t}{S}$ and $\textsc{w}_{\weightsFun}(f^{v \rightarrow c(v)}) = \weightsFun(v) \weightsSum \textsc{w}_{\weightsFun^{-v}}(f)$.
Furthermore, it is also straightforward to prove that for every $(X_t, c, \eta, \checkFuns)$--coloring $g$ in $\removeEdgesSet{G_t}{S}$, the function $g^{-v}$ is a $(X_s, c^{-v}, \eta^{\sim v}, \checkFuns^{-v})$--coloring in $\removeEdgesSet{G_s}{(S - \{v\})}$ and $\weightsFun(v) \weightsSum \textsc{w}_{\weightsFun^{-v}}(g^{-v}) = \textsc{w}_{\weightsFun}(g)$.
Therefore
\begin{align*}
	&\genDomSym_t(S, c, \weightsFun, \eta, \checkFuns)\\
=\;	&\minWeights\setst{\textsc{w}_{\weightsFun}(g)}{\text{$g$ is a $(X_t, c, \eta, \checkFuns)$--coloring in $\removeEdgesSet{G_t}{S}$}}\\
=\;	&\minWeights\setst{\weightsFun(v) \weightsSum \textsc{w}_{\weightsFun^{-v}}(f)}{\text{$f$ is a $(X_s, c^{-v}, \eta^{\sim v}, \checkFuns^{-v})$--coloring in $\removeEdgesSet{G_s}{(S - \{v\})}$}}\\
=\;	&\weightsFun(v) \weightsSum \minWeights\setst{\textsc{w}_{\weightsFun^{-v}}(f)}{\text{$f$ is a $(X_s, c^{-v}, \eta^{\sim v}, \checkFuns^{-v})$--coloring in $\removeEdgesSet{G_s}{(S - \{v\})}$}}\\
=\;	&\weightsFun(v) \weightsSum \genDomSym_s(S - \{v\}, c^{-v}, \weightsFun^{-v}, \eta^{\sim v}, \checkFuns^{-v}).
\end{align*}
\end{proof}
\end{lemma}

%-----------------------------
% Join
\begin{lemma}[Join node]\label{lem:join}
Let $t$ be a join node and $r$ and $s$ be the children of $t$.

We say that a pair $(\eta_r, \eta_s)$ of valid partial neighborhood mappings for $c$ is \emph{good} if $\checkFuns_v(\eta(v) \neighborhoodSum{v}{c(v)} \pnsubg(v,c,\removeEdgesSet{G_t}{S}[X_t]) \neighborhoodSum{v}{c(v)} \eta_r(v) \neighborhoodSum{v}{c(v)} \eta_s(v))$ is true for all $v \in X_t$.

Let $W = \bigweightsSum_{v \in X_t} \weightsFun(v)$. Then
$$\genDomSym_t(S, c, \weightsFun, \eta, \checkFuns) =
\minWeights_{(\eta_r, \eta_s) \text{ is good}}
\{
W \weightsSum \genDomSym_r(X_r, c, \weightsFun^{e}_{X_r}, \eta^e_c, \checkFuns^{\eta_r}) \weightsSum \genDomSym_s(X_s, c, \weightsFun^{e}_{X_s}, \eta^e_c, \checkFuns^{\eta_s})
\}.$$
\begin{proof}
Recall that $X_t = X_r = X_s$.

For every color assignment $f$ valid in $V(\removeEdgesSet{G_t}{S})$, denote by $f|_r$ (resp. $f|_s$) the restriction of $f$ to $V(\removeEdgesSet{G_r}{X_r})$ (resp. $V(\removeEdgesSet{G_s}{X_s})$).
Let $W = \bigweightsSum_{v \in X_t} \weightsFun(v)$.
Notice that $\textsc{w}_{\weightsFun}(f) = W \weightsSum \textsc{w}_{\weightsFun_{X_r}^{e}}(f|_r) \weightsSum \textsc{w}_{\weightsFun_{X_s}^{e}}(f|_s)$ for every color assignment $f$ valid in $V(\removeEdgesSet{G_t}{S})$.

Suppose there exists a $(X_t, c, \eta, \checkFuns)$--coloring in $\removeEdgesSet{G_t}{S}$ and let $f$ be one of them.
Let $\eta_r$ and $\eta_s$ be functions of domain $X_t$ such that $\eta_r(v) = \pnsubg(v, f, \removeEdgesSet{G_r}{X_r})$ and $\eta_s(v) = \pnsubg(v, f, \removeEdgesSet{G_s}{X_s})$ for all $v \in X_t$.

Since $V(\removeEdgesSet{G_r}{X_r}) \cap V(\removeEdgesSet{G_s}{X_s}) = X_t$ then $V(\removeEdgesSet{G_r}{X_r}) \cap V(\removeEdgesSet{G_s}{X_s}) \cap N_{\removeEdgesSet{G}{X_t}}(v) = \emptyset$. Moreover, since $f$ is a $(X_t, c, \eta, \checkFuns)$--coloring in $\removeEdgesSet{G_t}{S}$ then we know that, for all $v \in X_t$, $\pnsubg(v,c,\removeEdgesSet{G_t}{S}[X_t]) \neighborhoodSum{v}{c(v)} \eta_r(v) \neighborhoodSum{v}{c(v)} \eta_s(v) = \pnsubg(v, f, \removeEdgesSet{G_t}{S})$ and thus $\checkFuns_{v}(\eta(v) \neighborhoodSum{v}{c(v)} \pnsubg(v,c,\removeEdgesSet{G_t}{S}[X_t]) \neighborhoodSum{v}{c(v)} \eta_r(v) \neighborhoodSum{v}{c(v)} \eta_s(v)) = \True$.
Therefore, the functions $\eta_r$ and $\eta_s$ form a good pair of valid partial neighborhood mappings for $c$.

It is straightforward to prove that $f|_r$ is a $(X_r, c, \eta^e_c, \checkFuns^{\eta_r})$--coloring in $\removeEdgesSet{G_r}{X_r}$, and that $f|_s$ is a $(X_s, c, \eta^e_c, \checkFuns^{\eta_s})$--coloring in $\removeEdgesSet{G_s}{X_s}$.
Hence,
\begin{align*}
\textsc{w}_{\weightsFun}(f) &= W \weightsSum \textsc{w}_{\weightsFun_{X_r}^{e}}(f|_r) \weightsSum \textsc{w}_{\weightsFun_{X_s}^{e}}(f|_s)\\
	&\geq W \weightsSum \genDomSym_r(X_r, c, \weightsFun_{X_r}^{e}, \eta^e_c, \checkFuns^{\eta_r}) \weightsSum \genDomSym_s(X_s, c, \weightsFun_{X_s}^{e}, \eta^e_c, \checkFuns^{\eta_s})\\
	&\geq \minWeights_{(\eta_r, \eta_s) \text{ is good}}\{ W \weightsSum \genDomSym_r(X_r, c, \weightsFun^{e}_{X_r}, \eta^e_c, \checkFuns^{\eta_r}) \weightsSum \genDomSym_s(X_s, c, \weightsFun^{e}_{X_s}, \eta^e_c, \checkFuns^{\eta_s}) \}.
\end{align*}

Since $\minWeights\setst{\textsc{w}_{\weightsFun}(f)}{\text{$f$ is a $(X_t, c, \eta, \checkFuns)$--coloring in $\removeEdgesSet{G_t}{S}$}} = \Error$ if there are no $(X_t, c, \eta, \checkFuns)$--colorings in $\removeEdgesSet{G_t}{S}$, we obtain
\begin{align*}
	& \genDomSym_t(S, c, \weightsFun, \eta, \checkFuns)\\
=\;	& \minWeights\setst{\textsc{w}_{\weightsFun}(f)}{\text{$f$ is a $(X_t, c, \eta, \checkFuns)$--coloring in $\removeEdgesSet{G_t}{S}$}}\\
\geq\; & \minWeights_{(\eta_r, \eta_s) \text{ is good}}\{ W \weightsSum \genDomSym_r(X_r, c, \weightsFun^{e}_{X_r}, \eta^e_c, \checkFuns^{\eta_r}) \weightsSum \genDomSym_s(X_s, c, \weightsFun^{e}_{X_s}, \eta^e_c, \checkFuns^{\eta_s}) \}.
\end{align*}

To conclude, we will show that the other inequality holds.

If $\minWeights_{(\eta_r, \eta_s) \text{ is good}}\{ W \weightsSum \genDomSym_r(X_r, c, \weightsFun^{e}_{X_r}, \eta^e_c, \checkFuns^{\eta_r}) \weightsSum \genDomSym_s(X_s, c, \weightsFun^{e}_{X_s}, \eta^e_c, \checkFuns^{\eta_s}) \} = \Error$ then the statement trivially holds.
Otherwise, the minimum is realized by a good pair $(\widehat{\eta}_r, \widehat{\eta}_s)$, and $\genDomSym_r(X_r, c, \weightsFun^{e}_{X_r}, \eta^e_c, \checkFuns^{\widehat{\eta}_r}) \neq \Error$ and $\genDomSym_s(X_s, c, \weightsFun^{e}_{X_s}, \eta^e_c, \checkFuns^{\widehat{\eta}_s}) \neq \Error$.
Therefore there exists a $(X_r, c,  \eta^e_c, \checkFuns^{\widehat{\eta}_r})$--coloring $\widehat{f_r}$ in $\removeEdgesSet{G_r}{X_r}$ and a $(X_s, c,  \eta^e_c, \checkFuns^{\widehat{\eta}_s})$--coloring $\widehat{f_s}$ in $\removeEdgesSet{G_s}{X_s}$
such that
$\textsc{w}_{\weightsFun_{X_r}^{e}}(\widehat{f_r}) = \genDomSym_r(X_r, c, \weightsFun^{e}_{X_r}, \eta^e_c, \checkFuns^{\widehat{\eta}_r})$ and $\textsc{w}_{\weightsFun_{X_s}^{e}}(\widehat{f_s}) = \genDomSym_s(X_s, c, \weightsFun^{e}_{X_s}, \eta^e_c, \checkFuns^{\widehat{\eta}_s})$.

Let $\widehat{f}$ be a function of domain $V(\removeEdgesSet{G_t}{S})$ such that
\begin{itemize}
\item $\widehat{f}(v) = c(v)$ for all $v \in X_t$,
\item $\widehat{f}(v) = \widehat{f_r}(v)$ for all $v \in V(\removeEdgesSet{G_r}{X_r}) - X_r$, and
\item $\widehat{f}(v) = \widehat{f_s}(v)$ for all $v \in V(\removeEdgesSet{G_s}{X_s}) - X_s$.
\end{itemize}

It is straightforward to prove that $\widehat{f}$ is a $(X_t, c, \eta, \checkFuns)$--coloring in $\removeEdgesSet{G_t}{S}$, and also that $\widehat{f_r}$ (resp. $\widehat{f_s}$) is the restriction of $\widehat{f}$ to $V(\removeEdgesSet{G_r}{X_r})$ (resp. $V(\removeEdgesSet{G_s}{X_s})$).
Therefore, 
\begin{align*}
		& \minWeights_{(\eta_r, \eta_s) \text{ is good}}\{ W \weightsSum \genDomSym_r(X_r, c, \weightsFun^{e}_{X_r}, \eta^e_c, \checkFuns^{\eta_r}) \weightsSum \genDomSym_s(X_s, c, \weightsFun^{e}_{X_s}, \eta^e_c, \checkFuns^{\eta_s}) \}\\
=\;		& W \weightsSum \genDomSym_r(X_r, c, \weightsFun^{e}_{X_r}, \eta^e_c, \checkFuns^{\widehat{\eta}_r}) \weightsSum \genDomSym_s(X_s, c, \weightsFun^{e}_{X_s}, \eta^e_c, \checkFuns^{\widehat{\eta}_s})\\
=\;		& W \weightsSum \textsc{w}_{\weightsFun_{X_r}^{e}}(\widehat{f_r}) \weightsSum \textsc{w}_{\weightsFun_{X_s}^{e}}(\widehat{f_s})\\
=\;		& \textsc{w}_{\weightsFun}(\widehat{f})\\
\geq\;	& \minWeights\setst{\textsc{w}_{\weightsFun}(f)}{\text{$f$ is a $(X_t, c, \eta, \checkFuns)$--coloring in $\removeEdgesSet{G_t}{S}$}}\\
=\;		& \genDomSym_t(S, c, \weightsFun, \eta, \checkFuns).
\end{align*}

Consequently,
$$\genDomSym_t(S, c, \weightsFun, \eta, \checkFuns) = \minWeights_{(\eta_r, \eta_s) \text{ is good}}\{ W \weightsSum \genDomSym_r(X_r, c, \weightsFun^{e}_{X_r}, \eta^e_c, \checkFuns^{\eta_r}) \weightsSum \genDomSym_s(X_s, c, \weightsFun^{e}_{X_s}, \eta^e_c, \checkFuns^{\eta_s}) \}.$$
\end{proof}
\end{lemma}

Then there is a simple algorithm to compute $\genDomSym$. Indeed, based on the recurrence given in lemmas \ref{lem:leaf}, \ref{lem:forget}, \ref{lem:introduce} and \ref{lem:join}, the algorithm is executed in a bottom-up fashion (that is, first for all the leaf nodes, then for their parents, and so on) by computing $\genDomSym_t(S, c, \weightsFun, \eta, \checkFuns) \in \WeightSet$ for every node $t$ and every $(S, c, \weightsFun, \eta, \checkFuns) \in \genDomDomain{t}$.
Finally, the result is obtained by finding the minimum among all $\genDomSym_r(\emptyset, c, \weightsFun, \eta^e_c, \checkFuns)$ such that
$r$ is the root of $T$,
$c$ is a color assignment valid in $X_r$,
$\weightsFun \colon X_r \to \WeightSet$ is such that $\weightsFun(v) = \textsc{w}_{v, c(v)}$ for all $v \in X_r$,
and $\checkFuns_v = check_{v,c(v)}$ for all $v \in X_r$.

%---------------------------------------------------------------------------------------------------
\subsection{Time complexity}
Let $k = \max\setst{|X_t|}{t \in V(T)}$, $\mathcal{N} = \max\setst{|\NeighborhoodSet{v}{i}|}{v \in V(G), i \in L_v}$ and $\mathcal{C} = \max\setst{|L_v|}{v \in V(G)}$. Let $t_{\checkFuns}$, $t_{\neighborhoodSumSym}$, $t_{\toNeighborhoodSym}$, $t_{\weightsSum}$, and $t_{\minWeights}$ be upper bounds for the executing time of all the functions $check_{v,i}$ and $eq_{n}$, $\neighborhoodSum{v}{i}$, $\toNeighborhood{v}{i}$, $\weightsSum$, and $\minWeights$, respectively. Other operations are assumed to run in constant time. In particular, we are assuming that we access $\textsc{w}_{v,i}$ in constant time.

Traversing the tree $T$ requires $O(|V(T)|)$ time.
In $O(k^2|V(T)| + k|E(G)|)$ time we can construct the adjacency matrices of all the graphs $G[X_t]$ with $t \in V(T)$ (by traversing $T$ top-down and computing $N_{G_t}(v) \cap X_t$ only for nodes $t$ that are the child of a forget nodes $s$ with $X_t - X_s = \{v\}$).
Also, in $O(k)$ time we can construct each of the necessary function extensions and restrictions, and in $O((t_{\neighborhoodSumSym} + t_{\toNeighborhoodSym})k)$ we can construct each of the necessary $\eta^{\sim v}$.

We analyze four separate cases, and the proof in each one of them is straightforward.
\begin{itemize}
%-----------------------------
% Leaf
\item Leaf node: $O(t_{\checkFuns})$

%-----------------------------
% Forget
\item Forget node: $O(k^2 + (k + t_{\minWeights}) \mathcal{C})$

%-----------------------------
% Introduce
\item Introduce node: $O((t_{\neighborhoodSumSym} + t_{\toNeighborhoodSym}) k + t_{\checkFuns} + t_{\weightsSum} + k^2)$

%-----------------------------
% Join
\item Join node: $O((t_{\neighborhoodSumSym} + t_{\toNeighborhoodSym}) k^2 + t_{\weightsSum}k + ((t_{\neighborhoodSumSym} + t_{\checkFuns}) k + t_{\weightsSum} + t_{\minWeights}) \mathcal{N}^{2k})$
\end{itemize}

Each one of them is computed for every possible tuple $(S, c, \weightsFun, \eta, \checkFuns)$. We know that there are no more than $2^k \cdot \mathcal{C}^k \cdot 2^k \cdot \mathcal{N}^k \cdot (1 + \mathcal{N})^k$ of such tuples, and that constructing each of them requires $O(k)$ operations.

In summary, the time complexity of this algorithm is
$O(
(
	t_{\weightsSum}k
	+ (k + t_{\minWeights}) \mathcal{C}
	+ (t_{\neighborhoodSumSym} + t_{\toNeighborhoodSym}) k^2
	+ ((t_{\neighborhoodSumSym} + t_{\checkFuns}) k + t_{\weightsSum} + t_{\minWeights}) \mathcal{N}^{2k}
)
4^k \mathcal{C}^k \mathcal{N}^k (\mathcal{N} + 1)^k k
|V(T)|
+ k|E(G)|)$.

%---------------------------------------------------------------------------------------------------
\subsection{Special cases}
The next result easily follows from the previous time complexity analysis, Proposition~\ref{prop:twedges}, Theorem~\ref{thm:treedec} and Theorem~\ref{thm:nicetreedec}.

\begin{theorem}\label{thm:mainTheorem}
\THMmainTheorem
\end{theorem}

%---------------------------------------------------------------------------------------------------------
In particular, this result, together with Table~\ref{tab:partialNeighborhoods}, implies that all the problems listed in Table~\ref{tab:examples} can be solved in $O(|V(G)|)$ time.
For other problems, under certain hypothesis we can give a more generic {\pnPolynomial} partial neighborhood system, hence the next result.

\begin{corollary}\label{col:mainCorollary}
\COLmainCorollary

\begin{proof}
For each instance, let $\textsc{Colors} = \bigcup_{v \in V(G)} L_v$ and define the following partial neighborhood system:
\begin{itemize}
\item $\NeighborhoodSet{v}{i} = \intInterval{0}{d_G(v)}^{|\textsc{Colors}|}$;

\item $(n \neighborhoodSum{v}{i} n')_j = \min(n_j + n'_j, d_G(v))$ for all $j \in \textsc{Colors}$;

\item $\toNeighborhood{v}{i}(u,j)_j = 1$ and\\
$\toNeighborhood{v}{i}(u,j)_h = 0$ for all $h \in \textsc{Colors} - \{j\}$;

\item $check_{v,i}(n) = check(v,c)$ where $c$ is any color assignment valid in $N_G[v]$ such that $c(v) = i$ and $|\setst{u}{u \in N_G(v) \land c(u) = j}| = n_j$ for all $j \in \textsc{Colors}$. Note that we can construct $c$ in polynomial time using flow algorithms.
\end{itemize}

By Theorem \ref{thm:mainTheorem}, the statement holds.
\end{proof}
\end{corollary}

%-------------------------------------
\subsection{{\LCVP} problems}
In Section~\ref{sec:problemDescriptionLCVP} we have seen how to model the problem of deciding if $G$ has a $D_q$ partition as a {\rLCproblem{1}}, and now we extend it with a partial neighborhood system.
Let $m(S)$ be the maximum of $S$ if $S$ is finite, or the maximum of $\overline{S}$ if $S$ is co-finite. Then
\begin{itemize}
\item $\NeighborhoodSet{v}{i}$ is the Cartesian product of the sets $\intInterval{1}{m(D_q[i,j])+1}$ for all $1 \leq j \leq q$;

\item $n \neighborhoodSum{v}{i} n'$ is such that $(n \neighborhoodSum{v}{i} n')_j = \min(n_j + n'_j, m(D_q[i,j])+1)$ for all $1 \leq j \leq q$;

\item $\toNeighborhood{v}{i}(u, j)$ is the tuple such that its $j$th entry is 1 and all its other entries are 0; and

\item $check_{v,i}(n) = \left(\forallFormula{j \in \intInterval{1}{q}}{n_j \in D_q[i,j]}\right)$.
\end{itemize}

%Therefore, with the previous algorithm we can solve {\LCVP} problems in bounded treewidth graphs in $O( q^c |V(G)|)$ time, for some constant $c$.
%Therefore, with the previous algorithm we can solve {\LCVP} problems in bounded treewidth graphs in $O( q^{k+1} |V(G)|)$ time, where $k = \max\setst{|X_t|}{t \in V(T)}$ and $T$ is a tree-decomposition of $G$.
Therefore, with the algorithm in Section~\ref{subsec:alg}, we can solve {\LCVP} problems in bounded treewidth graphs in $O( q^c |V(G)|)$ time, for some constant $c$ (which equals $k+2$ if $k$ is the width of a given tree-decomposition of $G$).
%$O(
%(
%   t_{\weightsSum}k
%   + (k + t_{\minWeights}) \mathcal{C}
%   + (t_{\neighborhoodSumSym} + t_{\toNeighborhoodSym}) k^2
%   + ((t_{\neighborhoodSumSym} + t_{\checkFuns}) k + t_{\weightsSum} + t_{\minWeights}) \mathcal{N}^{2k}
%)
%4^k \mathcal{C}^k \mathcal{N}^k (\mathcal{N} + 1)^k k
%|V(T)|
%+ k|E(G)|)$.
% --> $O( (q k^2 + q k M^{2k}) 4^k q^k M^k (M + 1)^k k |V(T)| + k|E(G)|)$
%where $k = \max\setst{|X_t|}{t \in V(T)}$ and $M = \max\setst{m(D_q[i,j])+1}{1 \leq i,j \leq q}$.
This recovers the results obtained by Telle in~\cite{LCVSVP-thesis}.

\section{{\rLCproblems{r}} in bounded treewidth and bounded degree graphs}
\label{sec:boundedDegree}

%---------------------------------------------------------------------------------------------------
Recall the partial neighborhood system defined in Remark~\ref{vectorPartialNeighborhood}, for which $\mathcal{N} \leq (\mathcal{C}+2)^{\Delta(G)}$.
Hence, the next result easily follows from Theorem~\ref{thm:mainTheorem}.

\begin{corollary}\label{thm:polyBoundedTwDeg}
\THMpolyBoundedTwDeg
\end{corollary}

%---------------------------------------------------------------------------------------------------
Furthermore, the next lemma shows that fixed powers of bounded treewidth and bounded degree graphs are also bounded treewidth and bounded degree graphs, therefore extending the results of the previous sections to more problems in these graph classes.

\begin{lemma}\label{lem:powerBounds}
\LEMpowerBounds

\begin{proof}
The inequality $\Delta(G) \leq \Delta(G^p) \leq \Delta(G)^p$ follows easily from the definition of power of a graph. Let $v$ be a vertex of $G$ of maximum degree. In $G^p$, the graph induced by $N_{G}[v]$ is a clique of size $\Delta(G)+1$ and, by Theorem \ref{thm:twclique}, we get that $tw(G^p) \geq \Delta(G)$. Since $G$ is a subgraph of $G^p$ and by Proposition \ref{prop:twsubgraph}, we have $tw(G^p) \geq tw(G)$.

Now assume $(T, \{X_t\}_{t \in V(T)})$ is a tree decomposition of $G$. For every $t \in V(T)$, let $Y_t$ be the set of vertices that are at distance less than or equal to $\lceil\frac{p}{2}\rceil$ from a vertex of $X_t$.
We will prove that $(T, \{Y_t\}_{t \in V(T)})$, is a tree decomposition of $G^p$.
%-----------------------

Clearly, $\bigcup_{t \in V(T)} Y_t = V(G) = V(G^p)$, so property (W1) holds.
%-----------------------

Let $u, v$ be two vertices that are neighbors in $G^p$. If they are also neighbors in $G$, then there exists a bag $X_t$ that contains both of the vertices and since $X_t \subseteq Y_t$, we get that property (W2) holds in this case. If not, there exists a vertex $w$ that is at distance at most $\lceil\frac{p}{2}\rceil$ from both $u$ and $v$ in $G$. Therefore, since there exists a bag $X_t$ that contains $w$, this implies that $w \in Y_t$ (because $X_t \subseteq Y_t$) and $u,v \in Y_t$ (because $u,v$ are at distance at most $\lceil\frac{p}{2}\rceil$ from $w \in X_t$). Consequently, property (W2) also holds in this remaining case.
%-----------------------

Now we will prove that (W3) holds.
For all $u \in V(G)$, let $T^X_u = T[\setst{t \in V(T)}{u \in X_t}]$ and $T^Y_u = T[\setst{t \in V(T)}{u \in Y_t}]$. Applying property (W3) to $(T, \{X_t\}_{t \in V(T)})$, we obtain that $T^X_u$ is a subtree of $T$ for every $u \in V(G)$.
Let $v \in V(G)$. We will prove that $T^Y_v$ is connected.
By definition of the bags $Y_t$, we know that $v \in Y_t$ if and only if $t \in V(T^X_v)$ or $t \in V(T^X_u)$ for some $u$ such that $d(v,u) \leq \lceil\frac{p}{2}\rceil$.
Let $t \in V(T^X_v)$. Notice that in order to prove that $T^Y_v$ is connected it is sufficient to prove that there exists a path in $T$ between $t$ and every $s \in V(T^Y_v) - V(T^X_v)$. Let $s \in V(T^Y_v) - V(T^X_v)$ and let $v_s \in Y_s$ be such that $d(v,v_s) \leq \lceil\frac{p}{2}\rceil$.
Since $T^X_u$ is a subtree of $T$ for every $u \in V(G)$, and $V(T^X_u) \cap V(T^X_w) \neq \emptyset$ for all $uw \in E(G)$ (because of property (W2) applied to $(T, \{X_t\}_{t \in V(T)})$ and the edge $uw$), and there exists a path in $G$ between $v$ and $v_s$, we get that there exists a path in $T$ between $t$ and $s$.
Therefore (W3) holds for $(T, \{Y_t\}_{t \in V(T)})$.
%-----------------------

Since every bag $Y_t$ has at most $(tw(G)+1)(\Delta(G) + 1)^{\lceil\frac{p}{2}\rceil}$ vertices, we obtain $tw(G^p) \leq (tw(G)+1)(\Delta(G) + 1)^{\lceil\frac{p}{2}\rceil} - 1$.
\end{proof}
\end{lemma}

%---------------------------------------------------------------------------------------------------
Directly from Lemma~\ref{lem:powerBounds} and Corollary~\ref{thm:polyBoundedTwDeg}, by reducing the problem to a  {\rLCproblem{1}} in $G^{r}$, the next result easily follows.

\begin{corollary}\label{col:polyDistBoundedTwDeg}
\COLpolyDistBoundedTwDeg
\end{corollary}

As a result, the algorithm of Section~\ref{sec:boundedTreewidth} can be instantiated to solve, in polynomial time for bounded treewidth and bounded degree graphs,
distance coloring problems~\cite{L21-1996,L21-1992,CHANG200353,Lp1-2005,eccentric-color-trees,broadcast-chromatic,BRESAR20072303,FIALA2010771},
distance independence~\cite{distance-d-indep},
distance domination problems~\cite{FundamentalsDom}, and
distance LCVP problems~\cite{gen-dist-dom-mim-width-2019},
for bounded distances.

A similar result has been obtained by Jaffke, Kwon, Str{\o}mme and Telle for the distance versions of the LCVP problems in bounded MIM-width graphs~\cite{gen-dist-dom-mim-width-2019}.

\section{Dealing with global properties in bounded treewidth graphs}
\label{sec:globalProp}

In this section we explain how to modify the previous algorithm in order to handle some global properties.
The general idea in all of these cases is to modify $\genDomSym_t$ by extending it with new parameters (that is, at each node $t$ we compute $\genDomSym_t(S, c, \weightsFun, \eta, \checkFuns, \ldots)$).
For simplicity, in the following subsections we omit some parts of the original algorithm, writing only the necessary changes.

%---------------------------------------------------------------------------------------------------
\subsection{The size of a color class is an element of a particular set}
\label{subsec:globalProP:size}
Suppose we want the class of color $j$ to have a size that is an element of a set $\sigma \subseteq \mathbb{N}_0$.

Consider a deterministic finite-state automaton $(Q, \{1\}, \delta, q_0, F)$ that
accepts a string of $n$ consecutive characters $1$ if and only if $n \in \sigma$.
Notice that for all finite sets $\sigma \subseteq \mathbb{N}_0$ there exists such an automaton:
let $m$ be the maximum element of $\sigma$, $Q = \{s_0, \ldots, s_{m+1}\}$, $q_0 = s_0$, $F = \setst{s_i}{i \in \sigma}$, and $\delta(s_i, 1) = s_{i+1}$ for all $0 \leq i \leq m$ and $\delta(s_{m+1}, 1) = s_{m+1}$.
Although, for time complexity issues, when $m$ is not a constant we might be interested in another automata, with constant number of states (for example, if $\sigma$ is the set of odd numbers in $\intInterval{0}{|V(G)|}$, we only need two states).

In the algorithm, at each node $t$, we add a parameter $state_j$ that stores the state of the partial size of the color class $j$, and also a parameter $accept_j$ that checks if we are in the desired state, and then proceed in the following way.
\begin{itemize}[leftmargin=*]
%-----------------------------
% Leaf
\item \textbf{Leaf node:} Now we also need to check if $accept_j(state_j)$ is true.

%-----------------------------
% Forget
\item \textbf{Forget node:} For all $i \in L_v$, let $state_j^i = state_j$ if $i \neq j$ and $state_j^i = \delta(state_j, 1)$ otherwise. Then
$$\genDomSym_t(\ldots, state_j, accept_j) = \min\setst{\genDomSym_s(\ldots, state_j^i, accept_j)}{i \in L_v}.$$

%-----------------------------
% Introduce
\item \textbf{Introduce node:} Remains the same (with $state_j$ and $accept_j$ added to $\genDomSym_s$).

%-----------------------------
% Join
\item \textbf{Join node:}
For all $q \in Q$, let $eq_{q} \colon Q \to \BoolSet$ be such that $eq_{q}(q') = (q = q')$ for all $q' \in Q$. Then
$$\genDomSym_t(\ldots, state_j, accept_j) =
\minWeights\{W
\weightsSum \genDomSym_r(\ldots, state_j, eq_{q})
\weightsSum \genDomSym_s(\ldots, q, accept_j)
:
q \in Q \land
\ldots
\}.$$

%-----------------------------
% Root
\item At the root $r$ where $X_r = \{v\}$, we compute all $\genDomSym_r(\ldots, s_r, a)$, with $a$ such that $a(s) = (s \in F)$, and $s_r = q_0$ if $c(v) \neq j$ and $s_r = \delta(q_0, 1)$ otherwise.
\end{itemize}

Note that it is easy to generalize this idea to more classes by simply adding as many $state_j$ and $accept_j$ as needed (each of them with its own automaton), and even to a set $J$ of classes by replacing statements of the form ``$i \neq j$'' with ``$i \notin J$''.

The time complexity now depends on the number of states and color classes to restrict. We can assume that checking if a state is an accepting one is a constant-time operation and so is computing $\delta(s, 1)$. Let $\mathcal{R}$ be the number of color classes (or sets of color classes) to restrict and let $\mathcal{S}$ be the size of the largest set of states among all considered automata. The only changes in complexity are:
\begin{itemize}
\item \textbf{Leaf node:} add $\mathcal{R}$.
\item \textbf{Forget node:} add $2\mathcal{R}\mathcal{C}$.
\item \textbf{Introduce node:} add $2\mathcal{R}$.
\item \textbf{Join node:} multiply by $\mathcal{S}^{\mathcal{R}}$.
\item When we multiply by the number of all possible combinations of the parameters of $\genDomSym_t$: add a factor $(\mathcal{S} (\mathcal{S}+1))^{\mathcal{R}}$.
\end{itemize}
In particular, the complexity of the algorithm remains polynomial in $|V(G)|$ if $\mathcal{R}$ is bounded by a constant, allowing us to, for example, ask for a color class to be non-empty or to have at most one element.

%---------------------------------------------------------------------------------------------------
\subsection{{\LCC}-like properties}
\label{subsec:globalProP:monoids}
Let $(M, \oplus)$ be a commutative monoid.
Suppose we want to satisfy an expression of the form $\bigoplus_{v \in V(G)} f(v, c) \in X$ for some $X \subseteq M$ and function $f$ that receives a vertex $v$ and a color assignment valid in $N_G[v]$.
For all $V \subseteq V(G)$, let $M(V)$ be the set of different values that $\bigoplus_{v \in V} f(v, c)$ can have.
Assume that, for every $V \subseteq V(G)$, $|M(V)|$ is bounded by some polynomial in the size of the input. Also assume that computing $x \oplus y$ and $x \in X$ can be done in polynomial time for all $x,y \in M$ and $X \subseteq M$.

Let $(M^{*}, \oplus^{*})$ be obtained by adding an absorbing element $E^{*}$ to $(M, \oplus)$, and let $e^{*}$ be the neutral element.
Consider other items similar to the partial neighborhood system for $check$, but for $f$:
\begin{itemize}
\item A set $\widetilde{\NeighborhoodSet{v}{i}}$, for every $v \in V(G)$ and $i \in L_{v}$, together with a closed binary operation $\widetilde{\neighborhoodSumSym}^{v,i}$ on $\widetilde{\NeighborhoodSet{v}{i}}$ that is commutative and associative and has a neutral element $\widetilde{e}_{v,i}$.

\item A function $\widetilde{\toNeighborhood{v}{i}}$, for every $v \in V(G)$ and $i \in L_{v}$, that given $u \in N_G(v)$ and $j \in L_u$ returns an element of $\widetilde{\NeighborhoodSet{v}{i}}$.

\item A function $f_{v,i} \colon \widetilde{\NeighborhoodSet{v}{i}} \to M$, for every $v \in V(G)$ and $i \in L_{v}$. This function must satisfy
$f_{v,c(v)}\left(\widetilde{\bigneighborhoodSumSym}^{v,c(v)}_{u \in N_G(v)} \widetilde{\toNeighborhoodSym}_{v,c(v)}(u, c(u))\right) = f(v,c)$
for every vertex $v \in V(G)$ and every color assignment $c$ valid in $N_G[v]$.
\end{itemize}

In the algorithm, at each node $t$, we add to $\genDomSym_t$ the following parameters:
$x$, that stores the state of a partial accumulation of values $f(v, c)$;
$accept$, that checks if we are in the desired state (similar to the idea in the previous subsection);
$\widetilde{\eta}$, that carries the values of the ``partial neighborhood for $f$'' of the vertices in $X_t$; and
$\widetilde{f}$, that behaves like $\checkFuns$ except that instead of providing functions of codomain $\BoolSet$, it provides functions of codomain $M^{*}$.
Then proceed in the following way.
\begin{itemize}[leftmargin=*]
%-----------------------------
% Leaf
\item \textbf{Leaf node:} Now we also need to check if $accept(x \oplus^{*} \widetilde{f}_v(\widetilde{\eta}(v)))$ is true.

%-----------------------------
% Forget
\item \textbf{Forget node:}
$$\genDomSym_t(\ldots, x, accept, \widetilde{\eta}, \widetilde{f}) = \min\setst{\genDomSym_s(\ldots, x, accept, \widetilde{\eta}^{v \rightarrow \widetilde{e}_{v, i}}, \widetilde{f}^{v \rightarrow f_{v,i}})}{i \in L_v}.$$

%-----------------------------
% Introduce
\item \textbf{Introduce node:}
Accumulate in $x$ the value of $\widetilde{f}_v(\widetilde{\eta}(v))$, remove $v$ from the domain of the new functions, and compute the new partial neighborhoods of the vertices in $X_s$.

%-----------------------------
% Join
\item \textbf{Join node:}
For all $m \in M$, let $eq_{m} \colon M \to \BoolSet$ be such that $eq_{m}(m') = (m = m')$ for all $m' \in M$.
For all $\widetilde{\eta}$, let $\widetilde{f}^{\widetilde{\eta}}$ be the function defined as $\widetilde{f}^{\widetilde{\eta}}_v(n) = e^{*}$ if $n = \widetilde{\eta}(v)$ and $\widetilde{f}^{\widetilde{\eta}}_v(n) = E^{*}$ otherwise.
For all $X \subseteq V(G)$ and color assignment $c$ valid in $X$, let $\widetilde{\eta}_c^{\widetilde{e}}$ be the function of domain $X$ such that $\widetilde{\eta}_c^{\widetilde{e}}(v) = \widetilde{e}_{v, c(v)}$ for all $v \in X$.
For all $H$ subgraph of $G$ and all color assignment $c$ valid in $V(H)$, let $\widetilde{\pnsubg}(v,c,H) = \widetilde{\bigneighborhoodSumSym}^{v,c(v)}_{u \in N_{H}(v)} \widetilde{\toNeighborhoodSym}_{v,c(v)}(u, c(u))$ for all $v \in V(H)$.
Then

$\genDomSym_t(\ldots, x, accept, \widetilde{\eta}, \widetilde{f}) =
\minWeights\{ W \weightsSum
\genDomSym_r(\ldots, e, eq_{m_r}, \widetilde{\eta}_c^{\widetilde{e}}, \widetilde{f}^{\widetilde{\eta}_r})
\weightsSum
\genDomSym_s(\ldots, e, eq_{m_s}, \widetilde{\eta}_c^{\widetilde{e}}, \widetilde{f}^{\widetilde{\eta}_s})
:
m_r \in M(V(G_r) - X_r) \land
m_s \in M(V(G_s) - X_s) \land
accept(
    (\bigoplus^{*}_{v \in X_t}
    \widetilde{f}_v(
        \widetilde{\eta}(v) \widetilde{\neighborhoodSumSym}^{v,c(v)}
        \widetilde{\eta}_r(v) \widetilde{\neighborhoodSumSym}^{v,c(v)}\\
        \widetilde{\eta}_s(v)\widetilde{\neighborhoodSumSym}^{v,c(v)}
        \widetilde{\pnsubg}(v,c,\removeEdgesSet{G_t}{S}[X_t])
    ))
    \oplus^{*} x
)
\}.$

%-----------------------------
% Root
\item At the root $r$, with $X_r = \{v\}$, we compute all $\genDomSym_r(\ldots, e, a, \widetilde{\eta}_c^{\widetilde{e}}, \widetilde{f})$,
where
$a$ is such that $a(x) = (x \in X)$, and
$\widetilde{f}_v = f_{v,c(v)}$.
\end{itemize}

Note that it is easy to generalize this idea to more properties like this, by simply adding as many groups of parameters as needed. If the number of this properties is bounded by a constant, then the complexity of the algorithm remains polynomial.

This, along with the ideas and results in sections \ref{sec:problemDescriptionLCCECC}, \ref{sec:boundedTreewidth} and \ref{sec:boundedDegree}, recovers the main results from Bodlaender in~\cite{bodlaender1987,bodlaender1988}.

%---------------------------------------------------------------------------------------------------
\subsection{One color class is connected}
\label{subsec:globalProP:connected}
Suppose we want the class of color $j$ to be connected.

At each node $t$, we add the parameter $comp_j : X_t \to \intInterval{1}{k}$ that maps vertices of color $j$ to natural numbers that represent connected components.

In the following items, let $X_t^j$ denote $\setst{u}{u \in X_t \land c(u) = j}$, and $N_t^j(v)$ denote $N_{G}(v) \cap X_t^j$.

\begin{itemize}[leftmargin=*]
%-----------------------------
% Leaf
\item \textbf{Leaf node:} Remains the same.

%-----------------------------
% Forget
\item \textbf{Forget node:} $\genDomSym_t(\ldots, comp_j) = \min\setst{\genDomSym_s(\ldots, comp_j^i)}{i \in L_v}$
where $comp_j^i$ is such that:
\begin{itemize}
\item if $i \neq j$ then $comp_j^i(u) = comp_j(u)$ for all $u \in X_t^j$, and

\item $comp_j^j(v) = \begin{cases}
\min_{u \in N_t^j(v)}\{comp_j(u)\} & \text{if } N_t^j(v) \neq \emptyset \\
\text{any value in $\intInterval{1}{k} - \setst{comp_j(u)}{u \in X_t^j}$} & \text{otherwise}
\end{cases}$\\
and, for all $u \in X_t^j$, $comp_j^j(u) = comp_j^j(v)$ if there exists $z \in N_t^j(v)$ such that $comp_j(u) = comp_j(z)$, and $comp_j^j(u) = comp_j(u)$ otherwise.
\end{itemize}

In other words,  if $v$ is a neighbor of two or more vertices of different connected components then those connected components can be unified, and if $v$ is not a neighbor of any other vertex of color $j$ then it is in a new connected component.

%-----------------------------
% Introduce
\item \textbf{Introduce node:} If $c(v) \neq j$ then it remains the same (adding $comp_j$ to $\genDomSym_s$). Otherwise, we split the case related to $\checkFuns_v(n_v) = \True$ in the following ones:
	\begin{itemize}
	\item If there exists $u \in X_s^j$ such that $comp_j(v) = comp_j(u)$ then $\genDomSym_t(\ldots, comp_j) = \ldots \genDomSym_s(\ldots, comp_j^{-v})$.
	\item If there does not exist $u \in X_s^j$ such that $comp_j(v) = comp_j(u)$, and $X_s^j \neq \emptyset$ then $\genDomSym_t(\ldots, comp_j) = \Error$.
	\item If $X_s^j = \emptyset$ then let $M = (\{q_0, q_1\}, \{1\}, \delta, q_0, \{q_0\})$ be an automaton such that $\delta(q_0, 1) = q_1$ and $\delta(q_1, 1) = q_1$. We use $M$ to request that the class of color $j$ is empty in $G_s$, therefore $\genDomSym_t(\ldots, comp_j) = \ldots \genDomSym_s(\ldots, q_0, eq_{q_0})$.
	\end{itemize}

That is, if $v$ belongs to a different connected component than all the vertices of color $j$ in $X_s$, then there is no way to connect $v$ with them and we get an error. Also, if there are no vertices of color $j$ in $X_s$ then there cannot be any vertices of color $j$ in $G_s$, because $N_G(V(G) - V(G_s)) \cap V(G_s) \subseteq X_s$.

%-----------------------------
% Join
\item \textbf{Join node:} For every $S \subseteq \setst{comp_j(v)}{v \in X_t^j}$ let $comp_j^S$ be a function such that, for all $v \in X_t^j$,
$$comp_j^s(v) = \begin{cases}
\min(S)	&	\text{if } comp_j(v) \in S \\
comp_j(v)	&	\text{otherwise.}
\end{cases}$$

At this node we need to unify the different connected components. To do so, on one branch we unify a set $S$ of them while on the other branch we unify a set $R$ of them, such that $S \cup R = \setst{comp_j(v)}{v \in X_t^j}$ and $|S \cap R| = 1$. Then
$\genDomSym_t(\ldots, comp_j) =
\minWeights\{W
\weightsSum \genDomSym_r(\ldots, comp_j^S)
\weightsSum \genDomSym_s(\ldots, comp_j^R)
:
S \cup R = \setst{comp_j(v)}{v \in X_t^j} \land |S \cap R| = 1
\land \ldots
\}$.

%-----------------------------
% Root
\item At the root $r$ where $X_r = \{v\}$, the function $comp_j$ is such that $comp_j(v) = 1$.
\end{itemize}

%-----------------------------
As before, notice that it is easy to generalize this idea to more classes or sets of classes by adding as many $comp_j$ functions as needed.

Let $\mathcal{J}$ be the number of color classes (or sets of color classes) to restrict. The only changes in complexity are:
\begin{itemize}
\item \textbf{Leaf node:} add $\mathcal{J}$.
\item \textbf{Forget node:} add $(\mathcal{C}-1 + k^2)\mathcal{J}$.
\item \textbf{Introduce node:} add $k\mathcal{J}$.
\item \textbf{Join node:} multiply by $(k2^k)^{\mathcal{J}}$.
\item When we multiply by the number of all possible combinations of the parameters of $\genDomSym_t$: add a factor $(k^k)^{\mathcal{J}}$.
\end{itemize}

Note that because in the introduce node we require that the class of color $j$ is empty, we also need to compute $\genDomSym_t(\ldots, q_0, eq_{q_0})$ for all $t \in T$, but this addition does not change the time complexity here (due to the fact that both $\mathcal{S}$ and $\mathcal{R}$ are bounded by a constant in this case).

%---------------------------------------------------------------------------------------------------
\subsection{One color class is acyclic}
\label{subsec:globalProP:acyclic}
(For undirected graphs.)

It can be done in essentially the same way as for the connected property. The only difference is that in the introduce node we do as in the original algorithm, and in the forget node we check if $v$ is a neighbor of at least two vertices that belong to the same component (in which case there is a cycle and we raise an error).

\section{Applications}
\label{sec:examples}

In this section we show how to model different problems as {\rLCproblems{1}} with polynomial partial neighborhood systems. As a result, we obtain polynomial-time algorithms to solve these problems for bounded treewidth graphs. For double Roman domination, minimum chromatic violation and Grundy domination (and their variants), no such algorithms were previously known (until the date and to the best of our knowledge).
As regards the problems that were already known to be polynomial-time solvable for the before-mentioned classes, it is worth to mention how to restate these problems as {\rLCproblems{1}}, even when the time complexity of the proposed solution is worse than the best one known, because problems modeled this way can be easily modified or combined, adding global properties or more restrictions, or even dealing with some distance versions, and they can inspire the statement of other problems as {\rLCproblems{1}}.

Throughout this section, we will assume that, otherwise stated, the definitions
of $\NeighborhoodSet{v}{i}$ are for all $v \in V(G)$ and $i \in L_v$,
of $n \neighborhoodSum{v}{i} n'$ for all $v \in V(G)$, $i \in L_v$ and $n, n' \in \NeighborhoodSet{v}{i}$,
of $\toNeighborhood{v}{i}(u, j)$ for all $v \in V(G)$, $i \in L_v$, $u \in N_G(v)$ and $j \in L_u$,
and of $check_{v,i}(n)$ for all $v \in V(G)$, $i \in L_v$ and $n \in \NeighborhoodSet{v}{i}$.

%--------------------------------------------------------------------------------------------------------------------------------------------------
% New results
\subsection{Double Roman domination}
\label{sec:doubleRoman}

This problem was first defined in~\cite{doubleRoman} and proved to be NP-complete for bipartite and chordal graphs in~\cite{doubleRomanNPc}.

%-------------------------------------
A \emph{double Roman dominating function} on a graph $G$ is a function $f \colon V(G) \to \{0,1,2,3\}$ having the property that if $f(v) = 0$, then vertex $v$ must have at least two neighbors assigned 2 under $f$ or one neighbor $w$ with $f(w) = 3$, and if $f(v) = 1$, then vertex $v$ must have at least one neighbor $w$ with $f(w) \geq 2$. The \emph{weight} of a double Roman dominating function $f$ is the sum $f(V(G)) = \sum_{v \in V(G)} f(v)$, and the minimum weight of a double Roman dominating function on $G$ is the \emph{double Roman domination number} of $G$.

%-------------------------------------
We can model the Double Roman domination problem as a {\rLCproblem{1}} in the following way:
\begin{itemize}
\item $L_{v} = \{0,1,2,3\}$;

\item $(\WeightSet, \wlesseq, \weightsSum) = (\mathbb{R} \cup \{+\infty\}, \leq, +)$;

\item $\textsc{w}_{v, i} = i$;

\item $\begin{array}[t]{rrlrl}
check(v, c)
		& =		& (c(v) = 0	& \Rightarrow	& (\existsFormula{u,w \in N_G(v)}{u \neq w \land c(u) = c(w) = 2})\\
		&		&			& \lor			& (\existsFormula{u \in N_G(v)}{c(u) = 3}))\\
		& \land	& (c(v) = 1	& \Rightarrow	& \existsFormula{u \in N_G(v)}{c(u) \geq 2}).
\end{array}$
\end{itemize}

It is easy to see that some of its variants, such as perfect~\cite{perfectDoubleRoman}, independent~\cite{independentDoubleRoman}, outer independent~\cite{outerindepDoubleRoman} and total~\cite{totalDoubleRoman}, can be modeled by making slight modifications to the previous items.

By Corollary \ref{col:mainCorollary}, the Double Roman domination problem can be solved in polynomial time for graphs of bounded treewidth. However, we show a constant partial neighborhood system and therefore obtain a linear-time algorithm:
\begin{itemize}
\item $\NeighborhoodSet{v}{i} = \{0,1,2\} \times \{0, 1\}$;

\item $n \neighborhoodSum{v}{i} n' = (\min(n_1 + n'_1, 2), \min(n_2 + n'_2, 1))$;

\item $\toNeighborhood{v}{i}(u, j) = \begin{cases}
(0, 0) & \text{if } j \leq 1 \\
(1, 0) & \text{if } j = 2 \\
(0, 1) & \text{if } j = 3;
\end{cases}$

\item $check_{v,i}(n) = (i=0 \Rightarrow n_1 \geq 2 \lor n_2 \geq 1) \land (i=1 \Rightarrow n_1 + n_2 \geq 1)$.
\end{itemize}

Notice that this partial neighborhood system simply counts (until it saturates) the number of neighbors assigned with 2 and also the ones assigned with 3. For other versions of the double Roman domination problem, we might require to also count the number of neighbors assigned with 0 or 1.

\subsection{Minimum chromatic violation problem}
\label{sec:minimumChromaticViolation}

This NP-hard problem was first defined in \cite{minimum-chromatic-violation} as a generalization of the $k$-coloring problem.

%-------------------------------------
Given a graph $G$, a set of \emph{weak edges} $F \subseteq E(G)$ and a positive integer $k$, the \emph{minimum chromatic violation problem} asks for a $k$--coloring of the graph $G' = (V(G), E(G) - F)$ minimizing the number of weak edges with both endpoints receiving the same color.

%-------------------------------------
We can reduce this problem to the following {\rLCproblem{1}} in the subdivision graph $S(G)$:
%-------------------------------------
\begin{itemize}
\item $L_{v} = \intInterval{1}{k}$ for all $v \in V(G)$,\\
$L_{uv} = L_u \times L_v$ for all $uv \in E(G)$;

\item $(\WeightSet, \wlesseq, \weightsSum) = (\mathbb{R} \cup \{+\infty\}, \leq, +)$;

\item $\textsc{w}_{v,i} = 0$ for all $v \in V(G), i \in L_v$,\\
$\textsc{w}_{uv,(i,i)} = 1$ for all $uv \in E(G), i \in L_u \cap L_v$,\\
$\textsc{w}_{uv,(i,j)} = 0$ for all $uv \in E(G), i \in L_u, j \in L_v - \{i\}$;

\item $check(v, c) = \True$ for all $v \in V(G)$,\\
$check(uv, c) = (c(uv) = (c(u), c(v)))$ for all $uv \in F$, and\\
$check(uv, c) = (c(uv) = (c(u), c(v)) \land c(u) \neq c(v))$ for all $uv \in E(G) - F$.
\end{itemize}

Basically, every edge in $G$ is colored with a pair of colors and checks if these are the colors of its endpoints. Edges in $F$ are allowed to have endpoints of the same color, while edges not in $F$ always produce an error when colored with a pair of equal colors. If an edge is colored with a pair of equal colors then its weight is 1, otherwise is 0.

In this way we have $\mathcal{C} = k^2$.
We give a constant partial neighborhood system for this model:
\begin{itemize}
\item $\NeighborhoodSet{v}{i} = \BoolSet$ for all $v \in V(S(G)), i \in  L_v$;

\item $n \neighborhoodSum{v}{i} n' = (n \land n')$ for all $v \in V(S(G)), i \in  L_v$ and $n,n' \in \NeighborhoodSet{v}{i}$;

\item $\toNeighborhood{v}{i}(e,j) = \True$ for all $v \in V(G), i \in  L_v, e \in N_{S(G)}(v), j \in L_e$,\\
$\toNeighborhood{uv}{(c_u, c_v)}(u, j) = (j = c_u)$ for all $uv \in E(G), c_u \in  L_u, c_v \in L_v, j \in L_u$,\\
$\toNeighborhood{uv}{(c_u, c_v)}(v, j) = (j = c_v)$ for all $uv \in E(G), c_u \in  L_u, c_v \in L_v, j \in L_v$;

\item $check_{v,i}(n) = \True$ for all $v \in V(G), i \in L_v, n \in \NeighborhoodSet{v}{i}$,\\
$check_{uv,(i,j)}(n) = n$ for all $uv \in F, i \in L_u, j \in L_v, n \in \NeighborhoodSet{uv}{(i,j)}$,\\
$check_{uv,(i,j)}(n) = (n \land i \neq j)$ for all $uv \in E(G) - F, i \in L_u, j \in L_v, n \in \NeighborhoodSet{uv}{(i,j)}$.
\end{itemize}

Therefore, when $k$ is bounded by a constant the minimum chromatic violation problem can be solved in linear time for graphs of bounded treewidth.

\subsection{Grundy domination number}
\label{sec:grundy}

This problem was introduced in \cite{GrundyDom} and proved to be NP-complete even for chordal graphs.

%-------------------------------------
A sequence $S = \seq{v_1, \ldots, v_k}$ of distinct vertices of a graph $G$ is a \emph{dominating sequence} if $\{v_1, \ldots, v_k\}$ is a dominating set of $G$, and $S$ is called a \emph{legal (dominating) sequence} if (in addition) $N[v_i] - \bigcup_{j=1}^{i-1} N[v_j] \neq \emptyset$ for each $i$.
We say that $v_i$ \emph{footprints} the vertices in $N[v_i] - \bigcup_{j=1}^{i-1} N[v_j]$, and that $v_i$ is the \emph{footprinter} of every $u \in N[v_i] - \bigcup_{j=1}^{i-1} N[v_j]$ (notice that every vertex in $V(G)$ has a unique footprinter).
We are interested in the maximum length of a legal dominating sequence in $G$.

%-------------------------------------
Given a legal sequence $S$, every vertex $v \in V(G)$ can be associated with a pair $(p_v, f_v)$, where $p_v$ is the position of $v$ in $S$ (or $\notInSeq$ if $v$ is not in $S$) and $f_v$ is the position in $S$ of the footprinter of $v$.
Directly from the definition of legal sequences we can deduce the following statement. A set $\setst{(p_v, f_v)}{v \in V(G)}$ determines a legal sequence if and only if the following conditions are satisfied:
\begin{enumerate}
\item for all $v \in V(G)$, there exists a unique $u \in N_G[v]$ such that $f_v = p_u$, and 
 $f_v \leq p_w$ for all $w \in N_G[v]$ (i.e., $v$ is properly footprinted),
\item for all $v \in V(G)$, if $p_v \neq \notInSeq$ then there exists $u \in N_G[v]$ such that $f_u = p_v$ (i.e., if $v$ appears in the sequence then it footprints at least one vertex), and
\item $p_v \neq p_u$ for all $u,v \in V(G)$ such that $u \neq v$ and $p_u \neq \notInSeq$ (i.e., two vertices that appear in the sequence cannot have the same position in it).
\end{enumerate}

%-------------------------------------
Notice that conditions 1 and 2 are locally checkable, but the last one is not. However, we claim that the Grundy domination problem can be reduced to the following {\rLCproblem{1}}.
Let $n = |V(G)|$ and $\notInSeq = n+1$.
\begin{itemize}
\item $L_{v} = (\{1, \ldots, n\} \cup \{\notInSeq\}) \times \{1, \ldots, n\}$;

\item $(\WeightSet, \wlesseq, \weightsSum) = (\mathbb{R} \cup \{-\infty\}, \geq, +)$;

\item for all $f \in \{1, \ldots, n\}$, $\textsc{w}_{v, (\notInSeq, f)} = 0$ and $\textsc{w}_{v, (p, f)} = 1$ for all $p \neq \notInSeq$;

\item $\begin{array}[t]{rrl}
check(v, c)	& =		& (\existsUniqueFormula{u \in N_G[v]}{c(v)_2 = c(u)_1})\\
			& \land	& (\forallFormula{u \in N_G[v]}{c(v)_2 \leq c(u)_1})\\
			& \land	& (c(v)_1 \neq \notInSeq \Rightarrow \existsFormula{u \in N_G[v]}{c(v)_1 = c(u)_2})\\
			& \land	& (c(v)_1 \neq \notInSeq \Rightarrow \forallFormula{u \in N_G(v)}{c(v)_1 \neq c(u)_1}).
\end{array}$
\end{itemize}

%-------------------------------------
Let $G$ be a graph. Let $c$ be a proper coloring of $G$ in the previous {\rLCproblem{1}}, and let $(p_v, f_v) = c(v)$ for all $v \in V(G)$.
Conditions 1 and 2 are trivially satisfied.
Suppose that the last condition is not satisfied. Then there are two different vertices $u, v$ such that $p_u = p_v \neq \notInSeq$.
By definition of $check$, we know that
\begin{itemize}
\item $u \notin N_G[v]$,
\item if there exists a vertex $w$ such that $f_w > p_v = p_u$ then $v \notin N_G[w]$ and $u \notin N_G[w]$ (thus, $w \notin N_G[v]$ and $w \notin N_G[u]$), and
\item if there exists a vertex $w \in N_G[u]$ such that $f_w = p_v = p_u$ then $w \notin N_G[v]$.
\end{itemize}
Therefore, $N_G[u] \cap N_G[v] = F$ such that $f_w < p_u = p_v$ for all $w \in F$. Now we can assign colors $(p_w', f_w')$ for every $w \in V(G)$, such that
$$p_w' = \begin{cases}
p_w		& \text{if $p_w \leq p_u$ and $w \neq v$}\\
p_w + 1	& \text{otherwise}
\end{cases}
$$
and
$$f_w' = \begin{cases}
f_w		& \text{if $f_w < p_u$, or $f_w = p_u$ and $w \notin N_G[v]$}\\
f_w + 1	& \text{otherwise.}
\end{cases}
$$
That is, we move one position forward all the vertices that appear after $u$ in $S$ and increase the necessary $f_w$. It is easy to see that this new assignment preserves the ``legality'' of $u$ (i.e., if $N_G[u] - \setst{z}{z \in N_G[w] \land p_w < p_u} \neq \emptyset$ then $N_G[u] - \setst{z}{z \in N_G[w] \land p_w' < p_v'} \neq \emptyset$) and also of all the other vertices.

%-------------------------------------
In order to give a polynomial-time algorithm for the Grundy domination problem in bounded treewidth graphs, we define a constant partial neighborhood system as follows:
\begin{itemize}
\item $\NeighborhoodSet{v}{(p, f)} = \{0,1,2\} \times \BoolSet \times \BoolSet \times \BoolSet$;

\item $n \neighborhoodSum{v}{(p, f)} n' = (\min(n_1 + n_1', 2), n_2 \land n_2', n_3 \lor n_3', n_4 \land n_4')$;

\item $\toNeighborhood{v}{(p_v, f_v)}(u, (p_u, f_u)) = n$ where:
	\begin{itemize}
	\item $n_1 = \begin{cases}
	1	& \text{if $f_v = p_u$}\\
	0	& \text{otherwise,}
	\end{cases}$
	\item $n_2 = (f_v \leq p_u)$,
	\item $n_3 = (p_v = f_u)$,
	\item $n_4 = (p_v \neq p_u)$;
	\end{itemize}

\item $\begin{array}[t]{rrl}
check_{v,(p, f)}(n)
	& =		& (n_1 < 2 \land (n_1 = 1 \Rightarrow f \neq p) \land (n_1 = 0 \Rightarrow f = p))\\
	& \land	& (n_2 \land f \leq p)\\
	& \land	& (p \neq \notInSeq \Rightarrow (n_3 \lor p = f))\\
	& \land	& (p \neq \notInSeq \Rightarrow n_4).
\end{array}$
\end{itemize}

We can also model the Grundy total domination problem (defined in \cite{GrundyTotalDom}) in a very similar way, by simply removing the cases where a vertex can footprint itself.

For both problems, since $\mathcal{C}$ is $O(|V(G)|^2)$, and we defined a constant partial neighborhood system, the time complexity of the algorithm is polynomial in $|V(G)|$ for a graph $G$ in a family of bounded treewidth graphs.

%---------------------------------------------------------------------------------------------------
% Well known
\subsection{Additive coloring}
\label{sec:additiveColoring}
Let $\eta$ be an upper bound of the additive chromatic number. It was shown in \cite{luckyChoiceNumber} that the additive chromatic number in a graph $G$ is at most $\Delta(G)^2 - \Delta(G) + 1$.

To model the additive coloring problem as a {\rLCproblem{1}} with a partial neighborhood system, we define the colors to be pairs of integers $(n,s)$, where $n$ represents the number assigned to the vertex and $s$ the sum of the numbers assigned to its neighbors. Formally:

\begin{itemize}
\item $L_{v} = \intInterval{1}{\eta} \times \intInterval{1}{\Delta(G)\eta}$;

\item $(\WeightSet, \wlesseq, \weightsSum) = (\mathbb{R} \cup \{+\infty\}, \leq, \max)$;

\item $\textsc{w}_{v, i} = i_1$;

\item $check(v, c) = \left( \forallFormula{u \in N_G(v)}{c(u)_2 \neq c(v)_2} \right) \land \left( c(v)_2 = \sum_{u \in N_G(v)} c(u)_1 \right)$.
\end{itemize}

It is straightforward to derive a polynomial partial neighborhood system for this model. One possibility is the following:
\begin{itemize}
\item $\NeighborhoodSet{v}{i} = \BoolSet \times \intInterval{1}{\Delta(G)\eta+1}$;

\item $n \neighborhoodSum{v}{i} n' = (n_1 \land n'_1, \min(n_2 + n'_2, \Delta(G)\eta+1))$;

\item $\toNeighborhood{v}{i}(u,j) = (i_2 \neq j_2, j_1)$

\item $check_{v,i}(n) = n_1 \land (i_2 = n_2)$.
\end{itemize}

Then $\mathcal{C}$ is $O(\Delta(G)\eta^2)$ and $\mathcal{N}$ is $O(\Delta(G)\eta)$, implying that there exists a polynomial-time algorithm to compute $\eta(G)$ for graphs $G$ in a class of graphs of bounded treewidth. Another polynomial time algorithm was obtained by R. Grappe, L. N. Grippo, and M. Valencia-Pabon (personal communication).

\subsection{Distance domination problems}

These problems can be naturally expressed as {\rLCproblems{r}}, for some $r$ depending on the characteristics of the problem. Moreover, some of them are in fact {\rLCproblems{1}}.

We will start by showing how to model the distance $k$-domination problem as a {\rLCproblem{1}}. To do this, we restate the problem in the following way: vertices receive integers from $0$ to $k$ (that represent their distance to a vertex of the distance $k$-dominating set), and vertices with a number greater than $0$ must satisfy the condition of having a neighbor with the preceding number assigned. Then

\begin{itemize}
\item $L_{v} = \intInterval{0}{k}$;

\item $(\WeightSet, \wlesseq, \weightsSum) = (\mathbb{R} \cup \{+\infty\}, \leq, +)$;

\item $\textsc{w}_{v, 0} = 1$ and\\
$\textsc{w}_{v, i} = 0$ for all $i \in \intInterval{1}{k}$;

\item $check(v, c) = \left(c(v) > 0 \Rightarrow \existsFormula{u \in N_G(v)}{c(u) = c(v) - 1} \right)$.
\end{itemize}

It only remains to show a constant partial neighborhood system in order to get a linear-time algorithm for this problem in bounded treewidth graphs:
\begin{itemize}
\item $\NeighborhoodSet{v}{i} = \BoolSet$;

\item $n \neighborhoodSum{v}{i} n' = n \lor n'$;

\item $\toNeighborhood{v}{i}(u, j) = (j = i-1)$;

\item $check_{v,i}(n) = (i > 0 \Rightarrow n)$.
\end{itemize}

Notice that with a similar argument we can model similar problems involving more than two sets in the partition. The idea is to make the colors indicate how far the vertices are from every other color. For example, if we have to color the graph with $\{r, g, b\}$ in such a way that every vertex is at distance at most 3 from a vertex of color $r$ and at distance at most 2 from a vertex of color $g$, following this idea to model the problem as a {\rLCproblem{1}}, our set of colors is $\{r_{0,1}, r_{0,2}, g_{1,0}, g_{2,0}, g_{3,0}, b_{1,1}, b_{2,1}, b_{3,1}, b_{1,2}, b_{2,2}, b_{3,2}\}$.

However, when there are restrictions over the distance between vertices of the same color, the previous approach would not work. We will now explain how to model these problems when the required distance is 2.

Let us work with the semitotal domination problem. We first restate the problem in order to differentiate two possible situations for a vertex in $D$ (colors $D_1$ and $D_2$) and two possible situations for a vertex not in $D$ (colors $\overline{D}$ and $\overline{D}_{*}$).
Vertices of color $D_1$ represent those vertices in $D$ that have a neighbor in $D$, vertices of color $D_2$ represent those vertices in $D$ that are at distance 2 of another vertex in $D$, vertices of color $\overline{D}_{*}$ represent those vertices not in $D$ that are the nexus between two vertices in $D$, and vertices of color $\overline{D}$ represent all the remaining ones.
Then we can set
\begin{itemize}
\item $L_{v} = \{D_1, D_2, \overline{D}, \overline{D}_{*}\}$;

\item $(\WeightSet, \wlesseq, \weightsSum) = (\mathbb{R} \cup \{+\infty\}, \leq, +)$;

\item $\textsc{w}_{v, D_1} = \textsc{w}_{v, D_2} = 1$, and\\
$\textsc{w}_{v, \overline{D}} = \textsc{w}_{v, \overline{D}_{*}} = 0$;

\item $\begin{aligned}[t]
check(v, c) =& \; (c(v) \in \{\overline{D}, D_1\} \Rightarrow \existsFormula{u \in N_G(v)}{c(u) \in \{D_1, D_2\}})\\
           \land& \; (c(v) = D_2 \Rightarrow \existsFormula{u \in N_G(v)}{c(u) = \overline{D}_{*}})\\
           \land& \; (c(v) = \overline{D}_{*} \Rightarrow \existsFormula{u,w \in N_G(v)}{u \neq w \land c(u), c(w) \in \{D_1, D_2\}}).
\end{aligned}$
\end{itemize}
And we can naturally give a constant partial neighborhood system:
\begin{itemize}
\item $\NeighborhoodSet{v}{i} = \{0,1,2\}$;

\item $n \neighborhoodSum{v}{i} n' = \min(n + n', 2)$;

\item If $i \in \{D_1, \overline{D}, \overline{D}_{*}\}$:\\
$\toNeighborhood{v}{i}(u, j) = \begin{cases}
1 & \text{if } j \in \{D_1, D_2\} \\
0 & \text{otherwise,}
\end{cases}$\\
if $i = D_2$:\\
$\toNeighborhood{v}{i}(u, j) = \begin{cases}
1 & \text{if } j = \overline{D}_{*} \\
0 & \text{otherwise;}
\end{cases}$

\item $check_{v,i}(n) = (n \geq 1)$ for $i \in \{\overline{D}, D_1, D_2\}$, and\\
$check_{v,\overline{D}_{*}}(n) = (n \geq 2)$.
\end{itemize}

Notice that in the restatement of semitotal domination we changed the ``duties'' of the vertices: the ones in charge of checking that a vertex of color $D_2$ is at distance $2$ of another vertex in $D$ are now the vertices of color $\overline{D}_{*}$.

Combining the ideas of the two previous problems we can model other related problems (such as total distance 2-domination) as {\rLCproblems{1}}.

\subsection{Problems involving edges}
Consider locally checkable problems where, for every edge, a certain condition comprising their endpoints and their consecutive edges must be satisfied.
We will show how to reduce these problems to {\rLCproblems{1}} in the jagged graph $J(G)$ of the input graph $G$.
We consider two kind of problems: when edges do not have requirements over other edges, and when they do.

For the first class of problems, the reduction is straightforward. We might need to duplicate the colors in order to distinguish the colors assigned to edges from the colors assigned to vertices.
As an example, we show how to model vertex cover:
\begin{itemize}
\item $L_{v} = \{0,1\}$ for all $v \in V(G)$ and\\
$L_{uv} = \{0\}$ for all $uv \in E(G)$;

\item $(\WeightSet, \wlesseq) = (\mathbb{R} \cup \{+\infty\}, \leq)$ and $\weightsSum = +$;

\item $\textsc{w}_{v', i} = i$ for all $v' \in V(J(G)), i \in L_{v'}$;

\item $check(v, c) = \True$ for all $v \in V(G)$, and\\
$check(uv, c) = (c(u) + c(v) \geq 1)$ for all $uv \in E(G)$.
\end{itemize}
And we give a constant partial neighborhood system:
\begin{itemize}
\item $\NeighborhoodSet{v'}{i} = \{0, 1\}$;

\item $n \neighborhoodSum{v'}{i} n' = \min(n + n', 1)$;

\item $\toNeighborhood{v'}{i}(u', j) = j$; and

\item $check_{v,i}(n) = \True$ for all $v \in V(G), i \in L_v$, and\\
$check_{uv,0}(n) = (n \geq 1)$ for all $uv \in E(G)$.
\end{itemize}

Edge cover is basically the same as vertex cover but interchanging vertices and edges.

As regards the second class of these problems, since two edges in $E(G)$ that share an endpoint are at distance two in $J(G)$, we can use the ideas from semitotal domination: the neighbors in $J(G)$ of the edges in $E(G)$ (which are vertices in $V(G)$) are the ones in charge of checking the requirements of the edges in $E(G)$.
We illustrate the maximum matching problem, for which we can set:
\begin{itemize}
\item $L_{v} = \{0\}$ for all $v \in V(G)$ and\\
$L_{uv} = \{0,1\}$ for all $uv \in E(G)$;

\item $(\WeightSet, \wlesseq) = (\mathbb{R} \cup \{-\infty\}, \geq)$ and $\weightsSum = +$;

\item $\textsc{w}_{v', i} = i$ for all $v' \in V(J(G)), i \in L_{v'}$;

\item $check(v, c) = \left(\sum_{u \in N_G(v)} c(uv) \leq 1\right)$ for all $v \in V(G)$, and\\
$check(uv, c) = \True$ for all $uv \in E(G)$.
\end{itemize}
And we give a constant partial neighborhood system:
\begin{itemize}
\item $\NeighborhoodSet{v'}{i} = \{0, 1\}$;

\item $n \neighborhoodSum{v'}{i} n' = \min(n + n', 1)$;

\item $\toNeighborhood{v'}{i}(u', j) = j$; and

\item $check_{v,0}(n) = (n \leq 1)$ for all $v \in V(G)$, and\\
$check_{uv,i}(n) = \True$ for all $uv \in E(G), i \in L_{uv}$.
\end{itemize}

\section*{Acknowledgements}
\label{sec:ack}

This work was partially supported by
ANPCyT PICT-2015-2218,
UBACyT Grants
	20020190100126BA,
	20020170100495BA and
	20020160100095BA
(Argentina),
and
Programa Regional MATHAMSUD MATH190013.
Carolina L. Gonzalez is partially supported by a CONICET doctoral fellowship.

\bibliographystyle{abbrv}

\bibliography{Z-References}

\newpage
\appendix
\section{Problems definitions}
\label{sec:problemsDef}

We define here the decision versions of the problems mentioned along the paper. Other similar problems can also be modeled as {\rLCproblems{r}}.

%---------------------------------------------------------------------------------------------------
% Domination
\subsection{Domination problems} $ $

\defProbEnglish
{Dominating set~\cite{ore1962theory}}
{A (weighted) graph $G$ and a positive integer $k$.}
{Does there exist $S \subseteq V(G)$ of size (weight) at most $k$ and such that $|N[v] \cap S| \geq 1$ for every $v \in V(G)$?}

\defProbEnglish
{Total domination~\cite{totalDomination}}
{A (weighted) graph $G$ and a positive integer $k$.}
{Does there exist $S \subseteq V(G)$ of size (weight) at most $k$ and such that $|N(v) \cap S| \geq 1$ for every $v \in V(G)$?}

\defProbEnglish
{$k$-tuple domination~\cite{doubleDomination}}
{A (weighted) graph $G$ and a positive integer $\ell$.}
{Does there exist $S \subseteq V(G)$ of size (weight) at most $\ell$ and such that $|N[v] \cap S| \geq k$ for every $v \in V(G)$?}

\defProbEnglish
{Total $k$-tuple domination~\cite{k-tuple-total-dom}}
{A (weighted) graph $G$ and a positive integer $\ell$.}
{Does there exist $S \subseteq V(G)$ of size (weight) at most $\ell$ and such that $|N(v) \cap S| \geq k$ for every $v \in V(G)$?}

\defProbEnglish
{$k$-domination~\cite{k-dom}}
{A (weighted) graph $G$ and a positive integer $\ell$.}
{Does there exist $S \subseteq V(G)$ of size (weight) at most $\ell$ and such that $|N(v) \cap S| \geq k$ for every $v \in V(G)\setminus S$?}

\defProbEnglish
{$\{k\}$-domination~\cite{complexity-k-dom-related}}
{A graph $G$ and a positive integer $\ell$.}
{Does there exist a function $f \colon V(G) \to \{0,1,\ldots,k\}$ of weight at most $\ell$ and such that $f(N[v]) \geq k$ for every $v \in V(G)$?}

\defProbEnglish
{$k$-rainbow domination~\cite{rainbowDomNPc}}
{A graph $G$ and a positive integer $\ell$.}
{Does there exist a function $f \colon V(G) \to 2^{\{1, \ldots, k\}}$ of weight ($\sum_{v \in V(G)} |f(v)|$) at most $\ell$ and such that for every vertex $v \in V(G)$ for which $f(v) = \emptyset$ we have $\bigcup_{u \in N_G[v]} f(u) = \{1, \ldots, k\}$?}

\defProbEnglish
{Semitotal dominating set~\cite{goddard2014semitotal}}
{A (weighted) graph $G$ with no isolated vertex and a positive integer $k$.}
{Does there exist a dominating set $D \subseteq V(G)$ of size (weight) at most $k$ and such that every vertex in $D$ is within distance two of another vertex in $D$?}

\defProbEnglish
{Distance $k$-domination \textnormal{(also called \textsc{$k$-covering})}~\cite{k-covering,distanceDomination}}
{A (weighted) graph $G$ and a positive integer $\ell$.}
{Does there exist a set $S \subseteq V(G)$ of size (weight) at most $\ell$ and such that every vertex in $G$ is within distance $k$ from a vertex in $S$?}

\defProbEnglish
{Connected dominating set~\cite{connectedDom}}
{A (weighted) graph $G$ and a positive integer $k$.}
{Does there exist a dominating set of $G$ of size (weight) at most $k$ that induces a connected subgraph of $G$?}

\defProbEnglish
{Roman domination~\cite{RomanDomination}}
{A graph $G$ and a positive integer $k$.}
{Does there exist a function $f \colon V(G) \to \{0,1,2\}$ of weight at most $k$ and such that every vertex $u \in V(G)$ for which $f(u)=0$ is adjacent to at least one vertex $v \in V(G)$ for which $f(v) = 2$?}

\defProbEnglish
{Grundy total domination~\cite{GrundyTotalDom}}
{A graph $G$ and a positive integer $k$.}
{Does there exist a sequence $\seq{v_1, \ldots, v_{\ell}}$ of distinct vertices of $G$ such that $\ell \geq k$, $\{v_1, \ldots, v_{\ell}\}$ is a dominating set of $G$ and $N(v_i) - \bigcup_{j=1}^{i-1} N(v_j) \neq \emptyset$ for each $i$?}

%---------------------------------------------------------------------------------------------------
% Coloring
\subsection{Coloring problems} $ $

\defProbEnglish
{$k$-coloring~\cite{ComputerIntract}}
{A graph $G$.}
{Does there exist a function $c \colon V(G) \to \{1, \ldots, k\}$ such that $c(u) \neq c(v)$ whenever $uv \in E(G)$?}

\defProbEnglish
{List-coloring~\cite{erdos1979choosability,vizing1976vertex}}
{A graph $G$ and a set $L(v)$ of colors for each vertex $v \in V(G)$.}
{Does there exist a proper coloring $c$ such that $c(v) \in L_v$ for all $v \in V(G)$?}

\defProbEnglish
{$k$-chromatic sum~\cite{sumColoring}}
{A graph $G$.}
{Is there a proper coloring $c$ of the graph $G$ such that $\sum_{v \in V} c(v) \leq k$?}

\defProbEnglish
{Additive coloring \textnormal{(also called \textsc{lucky labeling})}~\cite{luckyLabeling}}
{A graph $G$ and a positive integer $k$.}
{Does there exist a function $f \colon V(G) \to \{1, \ldots, k\}$ such that for every two adjacent vertices $u,v$ the sums of numbers assigned to their neighbors are different (that is, $\sum_{w \in N(u)} f(w) \neq \sum_{z \in N(v)} f(z)$)?}

\defProbEnglish
{Acyclic coloring~\cite{acyclic-coloring}}
{A graph $G$ and a positive integer $k$.}
{Does there exist a $k$-coloring of $G$ such that of every subgraph of $G$ spanned by vertices of two of the colors is acyclic (in other words, is a forest)?}

\defProbEnglish
{$L(h,k)$-labeling~\cite{Lhk-survey}}
{A graph $G$ and a positive integer $s$.}
{Does there exist a labeling of its vertices by integers between $0$ and $s$ such that adjacent vertices of $G$ are labeled using colors at least $h$ apart, and vertices having a common neighbor are labeled using colors at least $k$ apart?}

%---------------------------------------------------------------------------------------------------
% Independence
\subsection{Independence problems} $ $

\defProbEnglish
{Independent set~\cite{ComputerIntract}}
{A (weighted) graph $G$ and a positive integer $k$.}
{Does there exist an independent set of $G$ of size (weight) at least $k$?}

\defProbEnglish
{$k$-independent set \textnormal{(also called \textsc{distance $d$-independent set})}~\cite{k-independence,distance-d-indep}}
{A graph $G$ and a positive integer $s$.}
{Does there exist $X \subseteq V(G)$ of size at least $s$ such that the distance between every two vertices of $X$ is at least $k + 1$?}

%---------------------------------------------------------------------------------------------------
% Packing
\subsection{Packing problems} $ $

\defProbEnglish
{$\{k\}$-packing function~\cite{leoni2014}}
{A graph $G$ and a positive integer $\ell$.}
{Does there exist a function $f \colon V(G) \to \{0,1,\ldots,k\}$ of weight at least $\ell$ and such that $f(N[v]) \leq k$ for every $v \in V(G)$?}

\defProbEnglish
{$k$-limited packing~\cite{k-limited-packing}}
{A graph $G$ and a positive integer $\ell$.}
{Does there exist a function $f \colon V(G) \to \{0,1\}$ of weight at least $\ell$ and such that $f(N[v]) \leq k$ for every $v \in V(G)$?}

\defProbEnglish
{Packing chromatic number~\cite{BRESAR20072303}}
{A graph $G$ and a positive integer $k$.}
{Can $G$ be partitioned into disjoint classes $X_1, \ldots, X_k$ where vertices in $X_i$ have pairwise distance greater than $i$?}

%---------------------------------------------------------------------------------------------------
% Edges
\subsection{Problems involving edges} $ $

\defProbEnglish
{Matching~\cite{edmonds1965}}
{A(n) (edge weighted) graph $G$ and a positive integer $k$.}
{Does there exist a set $M \subseteq E(G)$ of pairwise non-adjacent edges of size (weight) at least $k$?}

\defProbEnglish
{Edge domination~\cite{edgeDominationNP}}
{A(n) (edge weighted) graph $G$ and a positive integer $k$.}
{Does there exist $F \subseteq E(G)$ of size (weight) at most $k$ and such that every edge in $E(G)$ shares an endpoint with at least one edge in $F$?}

\defProbEnglish
{Vertex cover~\cite{Karp1972}}
{A (weighted) graph $G$ and a positive integer $k$.}
{Does there exist $S \subseteq V(G)$ of size (weight) at most $k$ and such that every edge in $E(G)$ has at least one endpoint in $S$?}

\defProbEnglish
{Edge cover~\cite{ComputerIntract}}
{A(n) (edge weighted) graph $G$ and a positive integer $k$.}
{Does there exist $F \subseteq E(G)$ of size (weight) at most $k$ and such that every vertex in $V(G)$ belongs to at least one edge in $F$?}

\end{document}